\documentclass[floats,floatfix,showpacs,prl,twocolumn,superscriptaddress,nofootinbib,nolongbibliography,reprint]{revtex4-2}

\makeatletter
\let\@ORGREVTEXendnotemark\@endnotemark
\let\@ORGREVTEX@makefnmark@cite\@makefnmark@cite
\def\@endnotemark{\bgroup\@fileswfalse\@ORGREVTEXendnotemark\egroup}
\def\@makefnmark@cite{\bgroup\@fileswfalse\@ORGREVTEX@makefnmark@cite\egroup}
\makeatother

\usepackage{graphicx}
\usepackage{amssymb}
\usepackage{amsmath}
\usepackage{multirow}
\usepackage{longtable}
\usepackage{rotating}
\usepackage{float}
\usepackage{tabularx}
\usepackage{siunitx}
\DeclareSIUnit\solarmass{\ensuremath{M_\odot}}
\DeclareSIUnit\parsec{pc}
\DeclareSIUnit\deg{deg}
\DeclareSIUnit\year{yr}

\usepackage{prd_macros}

\usepackage[dvipsnames, usenames,table]{xcolor}

\definecolor{linkcolor}{rgb}{0.0,0.3,0.5}
\usepackage[unicode, colorlinks=true, linkcolor=linkcolor, citecolor=linkcolor, filecolor=linkcolor,urlcolor=linkcolor, pdfusetitle]{hyperref}

\def\arraystretch{0.7}

\newcommand{\LowMassMone}{36}
\newcommand{\LowMassMtwo}{29}
\newcommand{\HighMassMone}{85}
\newcommand{\HighMassMtwo}{65}
\newcommand{\LowMassAone}{0.13294973597140505}
\newcommand{\LowMassAtwo}{0.13839524245594084}
\newcommand{\HighMassAone}{0.7615995758526639}
\newcommand{\HighMassAtwo}{0.8545292202712563}
\newcommand{\LowMassElo}{0.0001}
\newcommand{\LowMassEhi}{0.0077}
\newcommand{\HighMassElo}{0.0001}
\newcommand{\HighMassEhi}{0.0316}
\newcommand{\LowMassZ}{0.03283374128737131}
\newcommand{\HighMassZ}{0.10887767193819738}

\newcommand{\HeavyChirpMassInjected}{64.59}
\newcommand{\HeavyTotalMassInjected}{150.00}
\newcommand{\HeavyMassRatioInjected}{0.76}

\newcommand{\HeavyMassOneInjected}{85.00}
\newcommand{\HeavyMassTwoInjected}{65.00}
\newcommand{\HeavyAOneInjected}{0.76}
\newcommand{\HeavyATwoInjected}{0.85}
\newcommand{\HeavyTiltOneInjected}{0.83}
\newcommand{\HeavyTiltTwoInjected}{1.85}

\newcommand{\HeavyPhiOneTwoInjected}{1.48}
\newcommand{\HeavyPhiJLInjected}{2.03}
\newcommand{\HeavyThetaJNInjected}{0.88}
\newcommand{\HeavyChiPInjected}{0.60}
\newcommand{\HeavyChiEffInjected}{0.19}
\newcommand{\HeavyLuminosityDistanceInjected}{521.1}
\newcommand{\HeavyRedshiftInjected}{0.11}
\newcommand{\HeavyIotaInjected}{0.68}
\newcommand{\HeavyRAInjected}{6.26}
\newcommand{\HeavyDECInjected}{-1.15}

\newcommand{\LightChirpMassInjected}{28.10}
\newcommand{\LightTotalMassInjected}{65.00}
\newcommand{\LightMassRatioInjected}{0.81}

\newcommand{\LightMassOneInjected}{36.00}
\newcommand{\LightMassTwoInjected}{29.00}
\newcommand{\LightAOneInjected}{0.13}
\newcommand{\LightATwoInjected}{0.14}
\newcommand{\LightTiltOneInjected}{1.53}
\newcommand{\LightTiltTwoInjected}{1.50}

\newcommand{\LightPhiOneTwoInjected}{5.52}
\newcommand{\LightPhiJLInjected}{0.77}
\newcommand{\LightThetaJNInjected}{2.31}
\newcommand{\LightChiPInjected}{0.13}
\newcommand{\LightChiEffInjected}{0.01}
\newcommand{\LightLuminosityDistanceInjected}{149.1}
\newcommand{\LightRedshiftInjected}{0.03}
\newcommand{\LightIotaInjected}{2.28}
\newcommand{\LightRAInjected}{1.50}
\newcommand{\LightDECInjected}{-1.24}

\newcommand{\ETDHeavyOptSNRRecovered}{968^{+2}_{-2}}
\newcommand{\ETDHeavyChirpMassRecovered}{64.59^{+0.05}_{-0.05}}
\newcommand{\ETDHeavyTotalMassRecovered}{150.00^{+0.11}_{-0.12}}
\newcommand{\ETDHeavyMassRatioRecovered}{0.76^{+0.01}_{-0.01}}
\newcommand{\ETDHeavyMassOneRecovered}{84.99^{+0.31}_{-0.32}}
\newcommand{\ETDHeavyMassTwoRecovered}{65.01^{+0.26}_{-0.26}}
\newcommand{\ETDHeavyAOneRecovered}{0.76^{+0.02}_{-0.02}}
\newcommand{\ETDHeavyATwoRecovered}{0.85^{+0.02}_{-0.02}}
\newcommand{\ETDHeavyTiltOneRecovered}{0.83^{+0.02}_{-0.02}}
\newcommand{\ETDHeavyTiltTwoRecovered}{1.85^{+0.03}_{-0.03}}
\newcommand{\ETDHeavyPhiOneTwoRecovered}{1.48^{+0.03}_{-0.03}}

\newcommand{\ETDHeavyChiPRecovered}{0.60^{+0.01}_{-0.01}}
\newcommand{\ETDHeavyChiEffRecovered}{0.188^{+0.003}_{-0.003}} 
\newcommand{\ETDHeavyLuminosityDistanceRecovered}{521^{+4}_{-4}}
\newcommand{\ETDHeavyRedshiftRecovered}{0.1090^{+0.0008}_{-0.0008}}

\newcommand{\CEHeavyOptSNRRecovered}{4372^{+2}_{-2}}
\newcommand{\CEHeavyChirpMassRecovered}{64.59^{+0.01}_{-0.01}}
\newcommand{\CEHeavyTotalMassRecovered}{150.00^{+0.03}_{-0.03}}
\newcommand{\CEHeavyMassRatioRecovered}{0.765^{+0.001}_{-0.001}}
\newcommand{\CEHeavyMassOneRecovered}{85.00^{+0.08}_{-0.07}}
\newcommand{\CEHeavyMassTwoRecovered}{65.00^{+0.06}_{-0.06}}
\newcommand{\CEHeavyAOneRecovered}{0.762^{+0.005}_{-0.004}}
\newcommand{\CEHeavyATwoRecovered}{0.854^{+0.004}_{-0.004}}
\newcommand{\CEHeavyTiltOneRecovered}{0.83^{+0.01}_{-0.01}}
\newcommand{\CEHeavyTiltTwoRecovered}{1.85^{+0.01}_{-0.01}}
\newcommand{\CEHeavyPhiOneTwoRecovered}{1.48^{+0.01}_{-0.01}}

\newcommand{\CEHeavyChiPRecovered}{0.604^{+0.003}_{-0.003}}
\newcommand{\CEHeavyChiEffRecovered}{0.1885^{+0.0008}_{-0.0008}}

\newcommand{\HLVHeavyOptSNRRecovered}{248^{+2}_{-2}}
\newcommand{\HLVHeavyChirpMassRecovered}{64.7^{+0.4}_{-0.4}}
\newcommand{\HLVHeavyTotalMassRecovered}{150.2^{+1.0}_{-0.9}}
\newcommand{\HLVHeavyMassRatioRecovered}{0.77^{+0.02}_{-0.03}}
\newcommand{\HLVHeavyMassOneRecovered}{85.0^{+1.5}_{-1.4}}
\newcommand{\HLVHeavyMassTwoRecovered}{65.2^{+1.1}_{-1.2}}
\newcommand{\HLVHeavyAOneRecovered}{0.7^{+0.2}_{-0.1}}
\newcommand{\HLVHeavyATwoRecovered}{0.7^{+0.2}_{-0.5}}
\newcommand{\HLVHeavyTiltOneRecovered}{0.8^{+0.3}_{-0.2}}
\newcommand{\HLVHeavyTiltTwoRecovered}{2.0^{+0.8}_{-0.2}}
\newcommand{\HLVHeavyPhiOneTwoRecovered}{1.3^{+4.5}_{-1.0}}
\newcommand{\HLVHeavyPhiJLRecovered}{1.9^{+0.4}_{-0.2}}
\newcommand{\HLVHeavyThetaJNRecovered}{0.91^{+0.06}_{-0.06}}
\newcommand{\HLVHeavyChiPRecovered}{0.6^{+0.2}_{-0.1}}
\newcommand{\HLVHeavyChiEffRecovered}{0.19^{+0.02}_{-0.02}}
\newcommand{\HLVHeavyLuminosityDistanceRecovered}{520^{+20}_{-20}}
\newcommand{\HLVHeavyRedshiftRecovered}{0.109^{+0.004}_{-0.005}}
\newcommand{\HLVHeavyIotaRecovered}{0.71^{+0.11}_{-0.08}}
\newcommand{\HLVHeavyRARecovered}{6.26^{+0.01}_{-0.01}}
\newcommand{\HLVHeavyDECRecovered}{-1.146^{+0.002}_{-0.002}}
\newcommand{\HLVHeavyCoalTimeRecovered}{-0.0001(3)}
\newcommand{\ETDLightOptSNRRecovered}{2701^{+2}_{-2}}
\newcommand{\ETDLightChirpMassRecovered}{28.096^{+0.001}_{-0.001}}
\newcommand{\ETDLightTotalMassRecovered}{65.00^{+0.01}_{-0.01}}
\newcommand{\ETDLightMassRatioRecovered}{0.805^{+0.003}_{-0.003}}
\newcommand{\ETDLightMassOneRecovered}{36.01^{+0.06}_{-0.06}}
\newcommand{\ETDLightMassTwoRecovered}{28.99^{+0.05}_{-0.05}}
\newcommand{\ETDLightAOneRecovered}{0.133^{+0.003}_{-0.003}}
\newcommand{\ETDLightATwoRecovered}{0.139^{+0.005}_{-0.005}}
\newcommand{\ETDLightTiltOneRecovered}{1.54^{+0.05}_{-0.05}}
\newcommand{\ETDLightTiltTwoRecovered}{1.50^{+0.07}_{-0.07}}

\newcommand{\ETDLightPhiOneTwoRecovered}{5.52^{+0.09}_{-0.09}}
\newcommand{\ETDLightPhiJLRecovered}{0.77^{+0.02}_{-0.02}}
\newcommand{\ETDLightThetaJNRecovered}{2.31^{+0.01}_{-0.01}}
\newcommand{\ETDLightChiPRecovered}{0.133^{+0.003}_{-0.003}}
\newcommand{\ETDLightChiEffRecovered}{0.0071^{+0.0005}_{-0.0005}}
\newcommand{\ETDLightLuminosityDistanceRecovered}{148.6^{+0.5}_{-0.4}}
\newcommand{\ETDLightRedshiftRecovered}{0.0328(1)}
\newcommand{\ETDLightIotaRecovered}{2.28^{+0.01}_{-0.01}}

\newcommand{\ETDLightCoalTimeRecovered}{0.00005^{+0.00015}_{-0.00015}}
\newcommand{\CELightOptSNRRecovered}{6837^{+3}_{-3}}
\newcommand{\CELightChirpMassRecovered}{28.0955^{+0.0006}_{-0.0006}}
\newcommand{\CELightTotalMassRecovered}{65.00^{+0.01}_{-0.01}}
\newcommand{\CELightMassRatioRecovered}{0.806^{+0.001}_{-0.001}}
\newcommand{\CELightMassOneRecovered}{36.00^{+0.03}_{-0.03}}
\newcommand{\CELightMassTwoRecovered}{29.00^{+0.02}_{-0.03}}
\newcommand{\CELightAOneRecovered}{0.133^{+0.002}_{-0.002}}
\newcommand{\CELightATwoRecovered}{0.138^{+0.002}_{-0.002}}
\newcommand{\CELightTiltOneRecovered}{1.53^{+0.03}_{-0.03}}
\newcommand{\CELightTiltTwoRecovered}{1.50^{+0.04}_{-0.04}}

\newcommand{\CELightPhiOneTwoRecovered}{5.52^{+0.05}_{-0.05}}
\newcommand{\CELightPhiJLRecovered}{0.78^{+0.01}_{-0.01}}
\newcommand{\CELightThetaJNRecovered}{2.312^{+0.005}_{-0.005}}
\newcommand{\CELightChiPRecovered}{0.133^{+0.002}_{-0.002}}
\newcommand{\CELightChiEffRecovered}{0.0070^{+0.0003}_{-0.0003}}
\newcommand{\CELightLuminosityDistanceRecovered}{52.03^{+0.03}_{-0.04}}
\newcommand{\CELightRedshiftRecovered}{0.011654(7)}
\newcommand{\CELightIotaRecovered}{2.275^{+0.005}_{-0.005}}

\newcommand{\HLVLightOptSNRRecovered}{401^{+2}_{-2}}
\newcommand{\HLVLightChirpMassRecovered}{28.10^{+0.04}_{-0.05}}
\newcommand{\HLVLightTotalMassRecovered}{65.01^{+0.12}_{-0.12}}
\newcommand{\HLVLightMassRatioRecovered}{0.81^{+0.01}_{-0.01}}
\newcommand{\HLVLightMassOneRecovered}{36.0^{+0.3}_{-0.3}}
\newcommand{\HLVLightMassTwoRecovered}{29.0^{+0.3}_{-0.3}}
\newcommand{\HLVLightAOneRecovered}{0.12^{+0.04}_{-0.04}}
\newcommand{\HLVLightATwoRecovered}{0.14^{+0.06}_{-0.06}}
\newcommand{\HLVLightTiltOneRecovered}{1.5^{+0.3}_{-0.3}}
\newcommand{\HLVLightTiltTwoRecovered}{1.5^{+0.4}_{-0.3}}

\newcommand{\HLVLightPhiOneTwoRecovered}{5.5^{+0.5}_{-0.6}}

\newcommand{\HLVLightThetaJNRecovered}{2.31^{+0.03}_{-0.02}}
\newcommand{\HLVLightChiPRecovered}{0.13^{+0.04}_{-0.04}}
\newcommand{\HLVLightChiEffRecovered}{0.01^{+0.01}_{-0.01}}
\newcommand{\HLVLightLuminosityDistanceRecovered}{149^{+3}_{-3}}
\newcommand{\HLVLightRedshiftRecovered}{0.0329(7)}
\newcommand{\HLVLightIotaRecovered}{2.28^{+0.07}_{-0.03}}
\newcommand{\HLVLightRARecovered}{1.495^{+0.002}_{-0.002}}
\newcommand{\HLVLightDECRecovered}{-1.239^{+0.001}_{-0.001}}
\newcommand{\HLVLightCoalTimeRecovered}{0.00000(6)}

\newcommand{\HLVLightOmega}{0.2}
\newcommand{\ETDLightOmega}{58.9}
\newcommand{\HLVHeavyOmega}{1.7}
\newcommand{\ETDHeavyOmega}{6.5}

\newcommand{\LISAHeavyLoEccSpinOneTilt}{0.92} 
\newcommand{\LISAHeavyLoEccSpinTwoTilt}{1.77} 
\newcommand{\LISALightLoEccSpinOneTilt}{1.45} 
\newcommand{\LISALightLoEccSpinTwoTilt}{1.60} 

\newcommand{\LISAHeavyLoEccPhiOneTwo}{2.95} 
\newcommand{\LISALightLoEccPhiOneTwo}{0.88} 

\newcommand{\LISAHeavyHiEccChieff}{0.188} 
\newcommand{\LISALightHiEccChieff}{0.0070} 

\newcommand{\LISAHeavyHiEccChip}{0.617} 
\newcommand{\LISALightHiEccChip}{0.132} 

\newcommand{\LISAHeavyHiEccIota}{0.542} 
\newcommand{\LISALightHiEccIota}{2.32} 

\newcommand{\LISAHeavyLoEccThetaJN}{0.700} 
\newcommand{\LISAHeavyHiEccThetaJN}{0.682} 
\newcommand{\LISALightLoEccThetaJN}{2.03} 
\newcommand{\LISALightHiEccThetaJN}{2.09} 

\newcommand{\LISAHeavyLoEccPhiJL}{0.216} 
\newcommand{\LISAHeavyHiEccPhiJL}{0.192} 
\newcommand{\LISALightLoEccPhiJL}{0.445} 
\newcommand{\LISALightHiEccPhiJL}{0.365} 

\newcommand{\LISAHeavyHiEccOmega}{10.7} 
\newcommand{\LISAHeavyLoEccOmega}{10.5} 
\newcommand{\LISALightHiEccOmega}{4.3} 
\newcommand{\LISALightLoEccOmega}{4.1} 

\newcommand{\HeavyHiEccInjected}{0.0316} 
\newcommand{\HeavyLoEccInjected}{0.0001} 
\newcommand{\LightHiEccInjected}{0.0077} 
\newcommand{\LightLoEccInjected}{0.0001} 

\newcommand{\HeavyLoChirpMassRecovered}{64.592^{+0.002}_{-0.001}} 
\newcommand{\HeavyLoTotalMassRecovered}{154^{+18}_{-6}} 
\newcommand{\HeavyLoMassRatioRecovered}{>0.38} 
\newcommand{\HeavyLoMassOneRecovered}{100^{+30}_{-20}} 
\newcommand{\HeavyLoMassTwoRecovered}{60^{+10}_{-10}} 
\newcommand{\HeavyLoAOneRecovered}{0.3^{+0.3}_{-0.2}} 
\newcommand{\HeavyLoATwoRecovered}{0.4^{+0.4}_{-0.4}} 
\newcommand{\HeavyLoTiltOneRecovered}{0.8^{+1.4}_{-0.5}} 
\newcommand{\HeavyLoTiltTwoRecovered}{2.0^{+0.8}_{-1.2}} 
\newcommand{\HeavyLoPhiOneTwoRecovered}{2.0^{+1.0}_{-2.0}} 
\newcommand{\HeavyLoChiPRecovered}{0.2^{+0.3}_{-0.1}}  
\newcommand{\HeavyLoChiEffRecovered}{0.22^{+0.05}_{-0.04}} 
\newcommand{\HeavyLoEccRecovered}{<0.0023} 
\newcommand{\HeavyLoLuminosityDistanceRecovered}{500.0^{+200.0}_{-100.0}}
\newcommand{\HeavyLoRedshiftRecovered}{0.10^{+0.03}_{-0.03}} 
\newcommand{\HeavyLoIotaRecovered}{0.9^{+0.3}_{-0.5}} 
\newcommand{\HeavyLoPhiJLRecovered}{0.2^{+0.3}_{-0.2}} 
\newcommand{\HeavyLoThetaJNRecovered}{0.9^{+0.4}_{-0.5}} 

\newcommand{\HeavyHiChirpMassRecovered}{64.594^{+0.001}_{-0.002}} 
\newcommand{\HeavyHiTotalMassRecovered}{151^{+14}_{-3}} 
\newcommand{\HeavyHiMassRatioRecovered}{>0.46} 
\newcommand{\HeavyHiMassOneRecovered}{90^{+30}_{-10}} 
\newcommand{\HeavyHiMassTwoRecovered}{60^{+10}_{-10}} 
\newcommand{\HeavyHiAOneRecovered}{0.3^{+0.3}_{-0.2}} 
\newcommand{\HeavyHiATwoRecovered}{0.3^{+0.5}_{-0.3}} 
\newcommand{\HeavyHiTiltOneRecovered}{0.8^{+1.3}_{-0.6}} 
\newcommand{\HeavyHiTiltTwoRecovered}{1.0^{+1.3}_{-0.7}} 
\newcommand{\HeavyHiPhiOneTwoRecovered}{2.0^{+1.0}_{-2.0}} 
\newcommand{\HeavyHiChiPRecovered}{0.2^{+0.3}_{-0.1}} 
\newcommand{\HeavyHiChiEffRecovered}{0.16^{+0.1}_{-0.07}} 
\newcommand{\HeavyHiEccRecovered}{0.0316^{+0.0002}_{-0.0001}} 
\newcommand{\HeavyHiLuminosityDistanceRecovered}{400^{+200}_{-100}} 
\newcommand{\HeavyHiRedshiftRecovered}{0.09^{+0.04}_{-0.02}} 
\newcommand{\HeavyHiIotaRecovered}{0.9^{+0.2}_{-0.5}} 
\newcommand{\HeavyHiPhiJLRecovered}{0.2^{+0.3}_{-0.2}} 
\newcommand{\HeavyHiThetaJNRecovered}{1.0^{+0.4}_{-0.5}} 

\newcommand{\LightHiChirpMassRecovered}{28.096^{+0.001}_{-0.001}} 
\newcommand{\LightHiTotalMassRecovered}{66^{+11}_{-2}} 
\newcommand{\LightHiMassRatioRecovered}{>0.36} 
\newcommand{\LightHiMassOneRecovered}{40^{+19}_{-8}} 
\newcommand{\LightHiMassTwoRecovered}{26^{+6}_{-7}} 
\newcommand{\LightHiAOneRecovered}{0.2^{+0.5}_{-0.2}} 
\newcommand{\LightHiATwoRecovered}{0.4^{+0.5}_{-0.3}} 
\newcommand{\LightHiTiltOneRecovered}{1.0^{+1.0}_{-1.0}} 
\newcommand{\LightHiTiltTwoRecovered}{1.0^{+1.0}_{-1.0}} 
\newcommand{\LightHiPhiOneTwoRecovered}{2.0^{+1.0}_{-2.0}} 
\newcommand{\LightHiChiPRecovered}{0.3^{+0.5}_{-0.2}} 
\newcommand{\LightHiChiEffRecovered}{0.0^{+0.2}_{-0.2}} 
\newcommand{\LightHiEccRecovered}{0.007^{+0.002}_{-0.002}}  
\newcommand{\LightHiLuminosityDistanceRecovered}{170^{+70}_{-60}} 
\newcommand{\LightHiRedshiftRecovered}{0.04^{+0.01}_{-0.01}} 
\newcommand{\LightHiIotaRecovered}{2.3^{+0.5}_{-0.4}} 
\newcommand{\LightHiPhiJLRecovered}{0.3^{+0.6}_{-0.3}} 
\newcommand{\LightHiThetaJNRecovered}{2.2^{+0.6}_{-0.6}} 

\newcommand{\LightLoChirpMassRecovered}{28.0954^{+0.0009}_{-0.0008}} 
\newcommand{\LightLoTotalMassRecovered}{68^{+13}_{-3}} 
\newcommand{\LightLoMassRatioRecovered}{>0.33} 
\newcommand{\LightLoMassOneRecovered}{40^{+20}_{-10}} 
\newcommand{\LightLoMassTwoRecovered}{24^{+7}_{-6}} 
\newcommand{\LightLoAOneRecovered}{0.3^{+0.5}_{-0.3}} 
\newcommand{\LightLoATwoRecovered}{0.4^{+0.5}_{-0.3}} 
\newcommand{\LightLoTiltOneRecovered}{1.1^{+1.1}_{-0.7}} 
\newcommand{\LightLoTiltTwoRecovered}{1.3^{+1.2}_{-0.9}} 
\newcommand{\LightLoPhiOneTwoRecovered}{2.0^{+1.0}_{-1.0}} 
\newcommand{\LightLoChiPRecovered}{0.3^{+0.4}_{-0.2}} 
\newcommand{\LightLoChiEffRecovered}{0.1^{+0.1}_{-0.1}} 
\newcommand{\LightLoEccRecovered}{<0.0047} 
\newcommand{\LightLoLuminosityDistanceRecovered}{170^{+70}_{-50}} 
\newcommand{\LightLoRedshiftRecovered}{0.04^{+0.02}_{-0.01}} 
\newcommand{\LightLoIotaRecovered}{2.3^{+0.5}_{-0.3}} 
\newcommand{\LightLoPhiJLRecovered}{0.3^{+0.5}_{-0.3}} 
\newcommand{\LightLoThetaJNRecovered}{2.2^{+0.6}_{-0.6}} 

\newcommand{\HeavyLoEccRARecovered}{6.26^{+0.02}_{-0.02}} 
\newcommand{\HeavyLoEccDECRecovered}{-1.144^{+0.007}_{-0.006}} 
\newcommand{\HeavyHiEccRARecovered}{6.27^{+0.01}_{-0.02}} 
\newcommand{\HeavyHiEccDECRecovered}{-1.147^{+0.008}_{-0.006}} 
\newcommand{\LightLoEccRARecovered}{1.50^{+0.01}_{-0.01}} 
\newcommand{\LightLoEccDECRecovered}{-1.24^{+0.004}_{-0.004}} 
\newcommand{\LightHiEccRARecovered}{1.50^{+0.01}_{-0.013}} 
\newcommand{\LightHiEccDECRecovered}{-1.239^{+0.004}_{-0.004}} 

\newcommand{\HeavyHiEccCoalTimeRecovered}{-2000^{+4000}_{-3000}} 
\newcommand{\HeavyLoEccCoalTimeRecovered}{0^{+2000}_{-4000}} 
\newcommand{\LightHiEccCoalTimeRecovered}{-1000^{+6000}_{-7000}} 
\newcommand{\LightLoEccCoalTimeRecovered}{1000^{+4000}_{-5000}} 

\newcommand{\HeavyLoEccSNRRecovered}{14^{+2}_{-2}} 
\newcommand{\HeavyHiEccSNRRecovered}{14^{+2}_{-2}} 
\newcommand{\LightLoEccSNRRecovered}{10^{+2}_{-2}} 
\newcommand{\LightHiEccSNRRecovered}{10^{+2}_{-2}} 

\newcommand{\HLV}{{HLV+{~}}}
\newcommand{\ecc}{{e_{0.01}}}
\newcommand{\mm}{\ensuremath{\mathrm{MM}}}

\makeatletter
\def\maketitle{\@author@finish \title@column\titleblock@produce \suppressfloats[t]}
\makeatother

\newcommand{\bham}{\affiliation{Institute for Gravitational Wave Astronomy \& School of Physics and Astronomy, University of Birmingham, Birmingham, B15 2TT, UK}}

\newcommand{\milan}{\affiliation{Dipartimento di Fisica ``G. Occhialini'', Universit\'a degli Studi di Milano-Bicocca, Piazza della Scienza 3, 20126 Milano, Italy} \affiliation{INFN, Sezione di Milano-Bicocca, Piazza della Scienza 3, 20126 Milano, Italy}}

\usepackage[dvipsnames]{xcolor}
\usepackage[normalem]{ulem}
\usepackage{soul}

\newcommand\orcidlink[1]{\href{https://orcid.org/#1}{$\!$\includegraphics[scale=0.006]{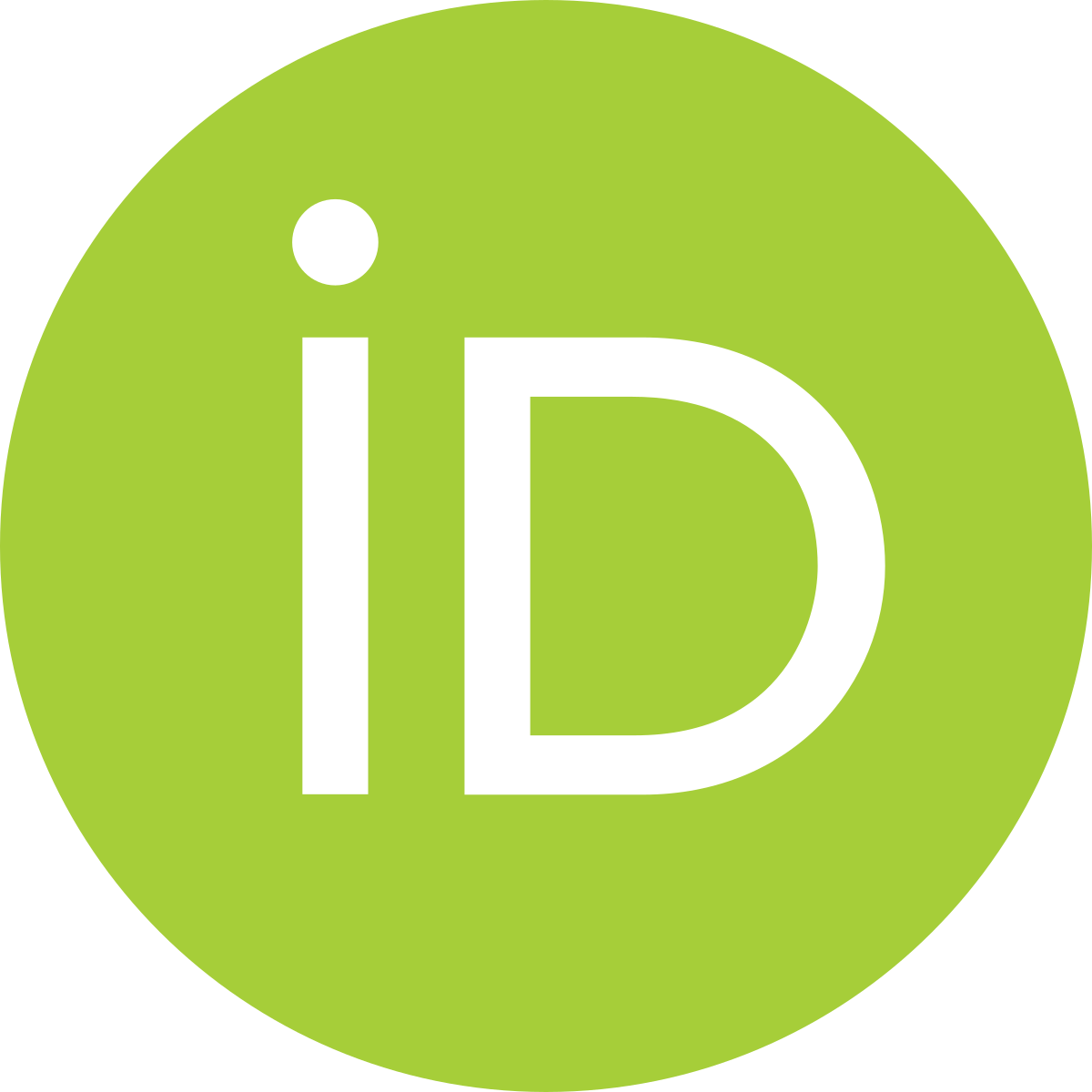} $\!\!$}}

\begin{document}

\title{The last three years: multiband
gravitational-wave observations of stellar-mass binary black holes}

\author{Antoine~Klein~\orcidlink{0000-0001-5438-9152}}\email{antoine@star.sr.bham.ac.uk }
\bham
\author{Geraint~Pratten}
\bham
\author{Riccardo~Buscicchio~\orcidlink{0000-0002-7387-6754}}
\bham \milan
\author{Patricia~Schmidt}
\bham
\author{Christopher~J.~Moore~\orcidlink{0000-0002-2527-0213}}
\bham
\author{Eliot~Finch~\orcidlink{0000-0002-1993-4263}}
\bham
\author{Alice~Bonino}
\bham
\author{Lucy~M.~Thomas}
\bham
\author{Natalie~Williams}
\bham
\author{Davide~Gerosa~\orcidlink{0000-0002-0933-3579}}
\bham \milan
\author{Sean~McGee~\orcidlink{0000-0003-3255-3139}}
\bham
\author{Matt~Nicholl}
\bham
\author{Alberto~Vecchio~\orcidlink{0000-0002-6254-1617}}
\bham

\date{\today}

\begin{abstract}
	Understanding the formation and evolution of the stellar-mass binary black holes discovered by LIGO and Virgo is a challenge that spans many areas of astrophysics, from stellar evolution, dynamics and accretion disks, to possible exotic early universe processes. 
	Over the final years of their lives, stellar-mass binaries radiate gravitational waves that are first observable by space-based detectors (such as LISA) and then ground-based instruments (such as LIGO, Virgo and the next generation observatories Cosmic Explorer and the Einstein Telescope). 
	Using state-of-the-art waveform models and parameter-estimation pipelines for both ground- and space-based observations, we show that (the expected handful of) these multiband observations will allow at least percent-level measurements of all 17 parameters that describe the binary, the possible identification of a likely host galaxy, and the forewarning of the merger days in advance allowing telescopes at multiple wavelengths to search for any electromagnetic signature associated to it.
	Multiband sources will therefore be a gold mine for astrophysics, but we also show that 
	they could be less useful as laboratories for fundamental tests of general relativity than has been previously suggested.
\end{abstract}

\maketitle

\noindent{\bf \em Introduction~--~}
The first detection of gravitational waves (GWs), GW150914~\cite{GW150914}, also revealed merging stellar-mass binary black holes (SmBBHs) with masses well above ${\sim 10\,M_\odot}$. 
Subsequently, tens of BBHs with component masses $\gtrsim 10\,M_\odot$ have been observed by the LIGO-Virgo network~\cite{GWTC3}.
The observed mass distribution of these black holes extends beyond the maximum mass predicted by the onset of (pulsational) pair-instability during the explosion of the progenitor star~\cite{Woosley:2007qp}, as firmly established by the detection of GW190521~\cite{GW190521}. 
No host galaxy of any BBH has so far been identified, although a possible counterpart of GW190521 has been suggested~\cite{Graham:2020gwr}. 
This means we have no direct information about the environment in which these systems form. 

Many potential formation and evolutionary scenarios for these BBHs have been suggested. These range from isolated formation to dynamical assembly in dense clusters or AGN disks, or less conventional routes such as primordial BH formation 
(see Refs.~\cite{2018arXiv180605820M,2021hgwa.bookE...4M,2021NatAs...5..749G} for reviews). 
Clearly identifying an evolutionary path is a holy grail of the field, with profound consequences for a range of open problems in stellar physics, binary (and higher multiplicity) stars, dynamics in dense environments, and relativistic astrophysics. 
Observationally, providing definitive evidence of a particular path is challenging, as it would ideally include a host galaxy identification and precise measurements of the orbital eccentricity and the masses and spins of the component BHs.

The progenitors to SmBBH mergers are potential sources~\cite{Sesana:2016ljz} for the Laser Interferometer Space Antenna (LISA)~\cite{LISA}.
LISA will detect SmBBHs radiating GWs at $\sim 10\,\mathrm{mHz}$ several years~\footnote{The title of this letter intentionally mirrors ``The last three minutes'' by  \citeauthor{1993PhRvL..70.2984C} \cite{1993PhRvL..70.2984C} --- a seminal study that set the stage for GW studies of compact binaries with ground-based detectors.} 
prior to merger. 
A handful of these systems are expected to be detected at low signal-to-noise ratios (SNRs) \cite{2019PhRvD..99j3004G,2019MNRAS.488L..94M,2022arXiv220102766S}.
These sources will become the first multiband GW signals when they merge $\sim 3\,$years later at $\sim 100\,\mathrm{Hz}$ in the sensitivity band of the ground-based instruments operating at the time (e.g.\ an improved version of the current LIGO-Virgo-KAGRA network or next generation (3G) observatories, such as Cosmic Explorer (CE)~\cite{Reitze:2019iox}
and/or the Einstein gravitational-wave Telescope (ET)~\cite{Punturo:2010zz}).

In this letter, we assess the scientific potential of these multiband sources using state-of-the-art waveforms and fully realistic Bayesian parameter estimation in both frequency regimes for the first time.
We consider two prototype sources with detector-frame (i.e.\ redshifted) BH masses $m_{1(2)}$ and dimensionless spins $\chi_{1(2)}$ consistent with those of GW150914 (``light'') and GW190521 (``heavy''), respectively. 
The minimum signal-to-noise ratio threshold needed for the detection of a SmBBH in LISA is uncertain; we simulate these systems with SNRs of $11.9$ and $15.3$ respectively which are close to threshold. 
For both systems, we simulate them from the start of LISA observations, which we take to be 3 years before merger, all the way to coalescence, when the GW signal is observed by ground-based observatories at SNR $\sim 10^3$.
We consider two possibilities for the orbital eccentricity (``low'' and ``high'') in the LISA band and assume that (due to GW emission) they have negligible eccentricity by the time they enter the frequency band of ground-based instruments \cite{Peters:1964zz}. 
The waveforms for these systems are shown in Fig.~\ref{fig:waveforms} and the system parameters are summarised in Table~\ref{tab:summary_parameters} (full details are given in the supplement).
We simulate these multiband sources consistently across all frequencies using waveforms that include eccentricity and spin-induced precession and higher multipoles where relevant.

\begin{table}
\caption{ \label{tab:summary_parameters}
	Summary of the properties of the injected sources. The orbital eccentricity at $10\,$mHz is denoted $\ecc$. 
} 
\begin{ruledtabular}
\begin{tabular}{l|cccccc}
  Name & $m_1$ & $m_2$ & $\chi_1$ & $\chi_2$ & $\ecc$ & $z$ \\
  \hline
  Light & 
  $\LowMassMone\,M_\odot$ & 
  $\LowMassMtwo\,M_\odot$ & 
  $\num[round-mode=places, round-precision=2]{\LowMassAone}$ & 
  $\num[round-mode=places, round-precision=2]{\LowMassAtwo}$ & 
  ${}^{\num[round-mode=figures, round-precision=3, scientific-notation=true]{\LowMassElo}}_{\num[round-mode=figures, round-precision=3, scientific-notation=true]{\LowMassEhi}}$ 
  &  \num[round-mode=figures, round-precision=2, scientific-notation=false]{\LowMassZ} \\
  Heavy & 
  $\HighMassMone\,M_\odot$ & 
  $\HighMassMtwo\,M_\odot$ & 
  \num[round-mode=places, round-precision=2]{\HighMassAone} & 
  \num[round-mode=places, round-precision=2]{\HighMassAtwo} & 
  ${}^{\num[round-mode=figures, round-precision=3, scientific-notation=true]{\HighMassElo}}_{\num[round-mode=figures, round-precision=3, scientific-notation=true]{\HighMassEhi}}$ 
  &  \num[round-mode=figures, round-precision=2, scientific-notation=false]{\HighMassZ} \\
\end{tabular}
\end{ruledtabular}
\end{table}

\begin{figure}[tbp]
	\centering
	\includegraphics[width=\columnwidth]{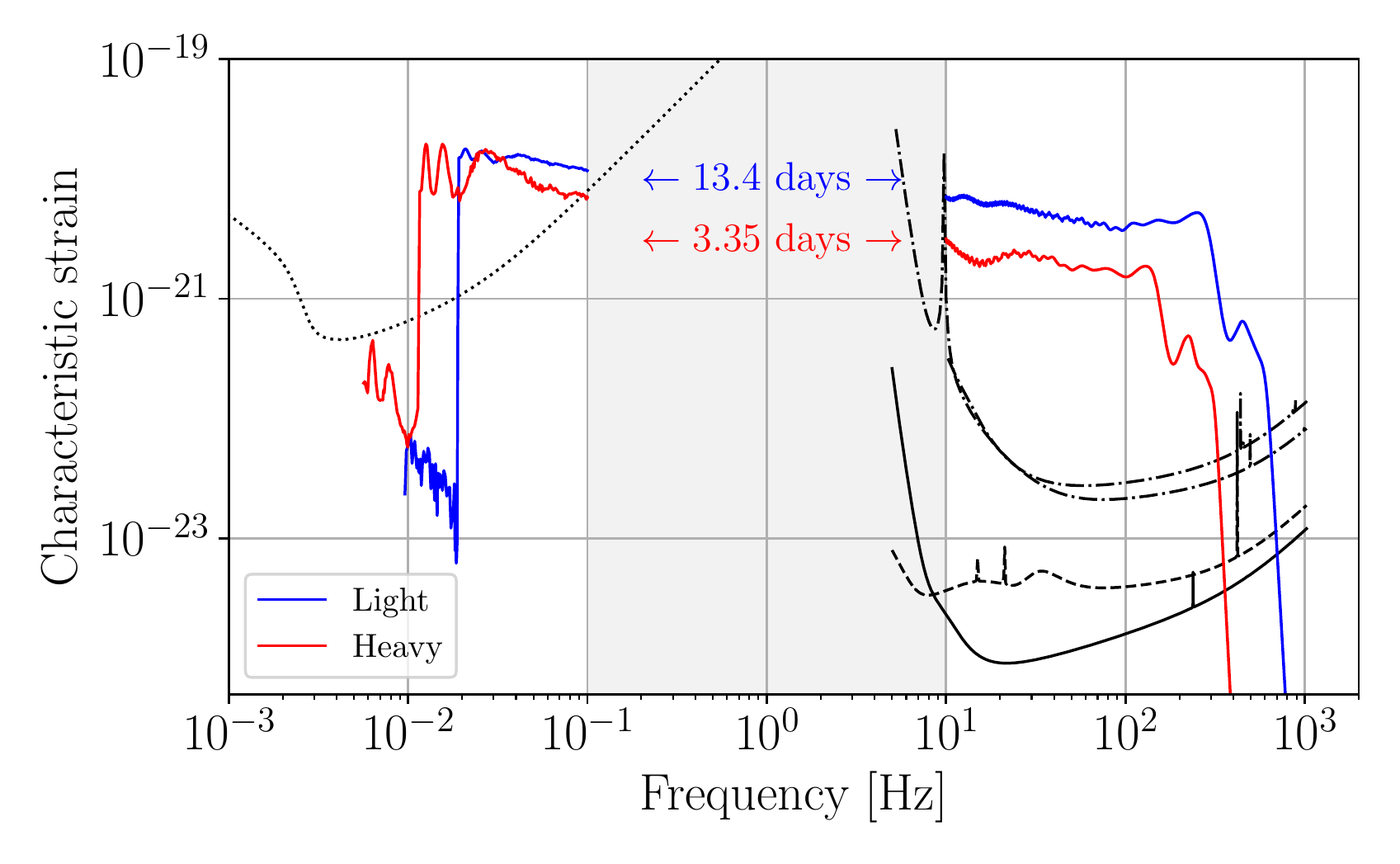}
	\caption{ \label{fig:waveforms}
		The multiband waveforms starting 3 years before merger chirp through the upper end of the LISA band at $\sim 10\,\mathrm{mHz}$ into the ground-based band (starting at $10\,\mathrm{Hz}$) where they merge. 
		Blue (red) solid lines show the characteristic strain for the light (heavy), high eccentricity sources.
		The sources take several days to cross from the LISA to the ground-based band.
		Black solid/dashed/dotted lines show the design sensitivities for CE/ET/LISA whilst the dash-dotted lines show the target O5 sensitivities of LIGO/Virgo.  
		Higher modes and spin precession characterize the signal modulation for ground based detectors.
		In the LISA band, signal modulation occurs due to spacecraft motion around the Sun and, for these high eccentricity sources, contributions from subdominant harmonics visible at the lowest frequencies.
		}
\end{figure}
%

\noindent{\bf \em LISA observations~--~}
The early inspiral is well described by post-Newtonian theory and is governed primarily by the chirp mass ${\cal M}_c = (m_1 m_2)^{3/5}/(m_1+m_2)^{1/5}$.
The inspiralling, precessing, and eccentric signal is modelled using an approach that leverages analytic solutions of the conservative problem to efficiently integrate the equations of motion over radiation-reaction timescales \cite{Klein:2021jtd}. 
Following \cite{balrog-ldc1}, we perform Bayesian inference on the full LISA time delay interferometry outputs using the \textsc{Balrog} code (for details, see the supplement). 

In all cases, the chirp mass is extremely well measured, with fractional errors of $\sim 10^{-4}$. Crucially, LISA also makes accurate measurements of the eccentricity, $\ecc$ (at a reference frequency of $10\,\mathrm{mHz}$), with absolute errors of $\sim 10^{-4} - 10^{-3}$.
The other intrinsic parameters (including the mass ratio and component spins) are not well measured, see figures and tables
in the supplement.

The LISA measurements of the chirp mass and eccentricity are partially degenerate with the time of merger, $t_m$, which is determined to within a few hours. 
When the system eventually merges in the ground-based band, $t_m$ is determined to within $\sim1\,\mathrm{ms}$, breaking this degeneracy.
Therefore, LISA measurements combined with just the time of the 3G detection (and no other information) allows a more precise determination of ${\cal M}_c$ and $\ecc$ (see Fig.~\ref{fig:Mc_e0_tm}).
The improved eccentricity measurements mean that binaries with residual eccentricities $\sim 10^{-3}$ at orbital periods of a few minutes can be distinguished from circular.

\begin{figure}[tbp]
	\centering
	\includegraphics[width=0.95\columnwidth]{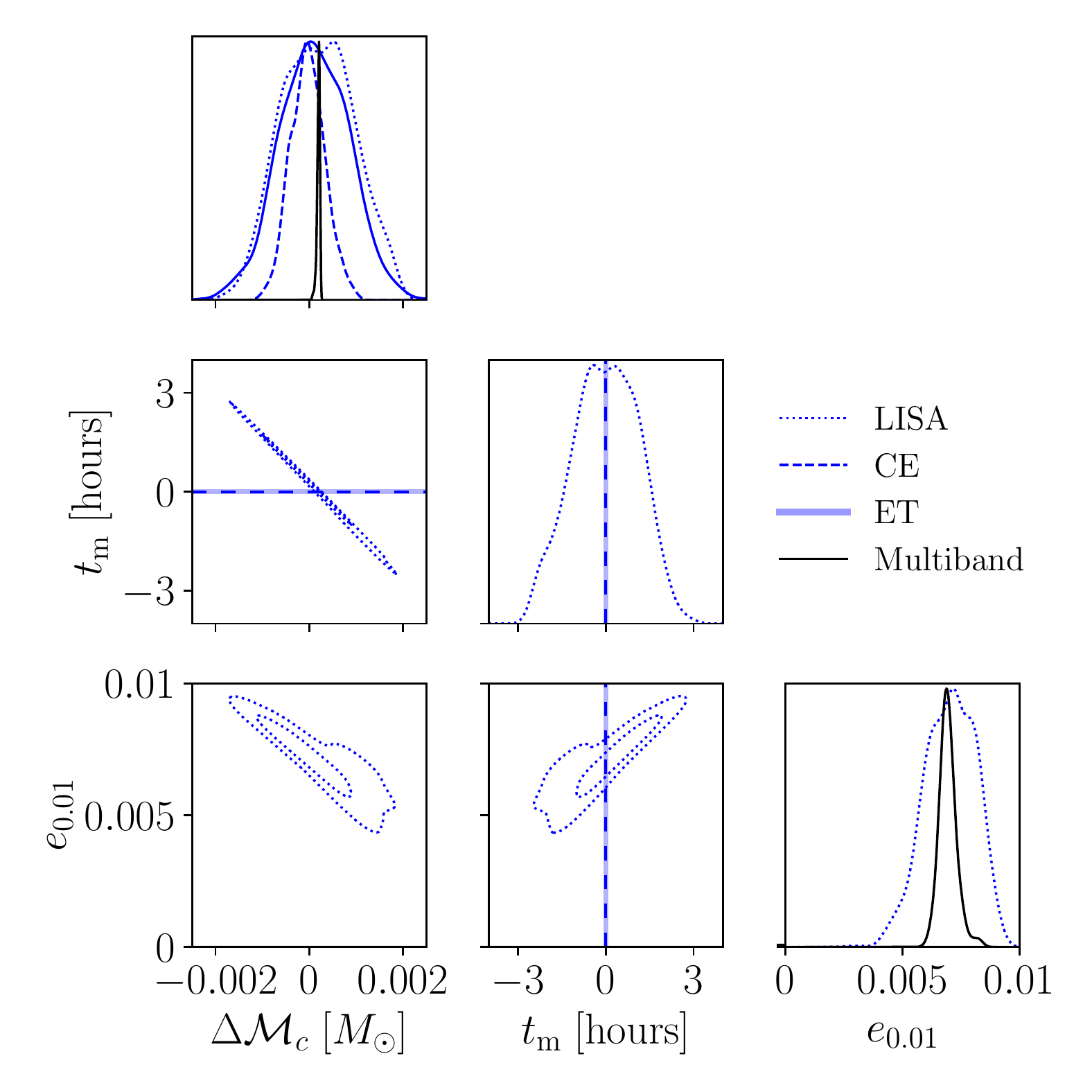}
	\caption{ \label{fig:Mc_e0_tm}
		LISA posteriors (dotted blue) on the chirp mass $\mathcal{M}_c$, time to merger $t_m$, and eccentricity $e_{0.01}$ for the ``light'',  high-eccentricity source.
		Solid/dashed blue lines show posteriors from third generation ET/CE ground-based instruments.
	    The LISA measurement of $\mathcal{M}_c$ (and, to a lesser extent, $e_{0.01}$) is degenerate with $t_m$ which is broken when combining with the merger time measured from any ground-based instrument leading to significant improvements in the combined, multiband measurements (shown in black). Similar results for the ``heavy'' source are shown in Fig.~\ref{fig:supp_Mc_tm_e0}.
	}
\end{figure}
%

\noindent{\bf \em Ground-based observations~--~}
It is currently uncertain what ground-based instruments will be operating when the first multiband sources merge.
We consider 3 possibilities: an upgrade (target O5 sensitivity) of the existing LIGO-Virgo network~\cite{psdobs} (hereafter \HLV), and the 3G instruments ET or CE.
Adopting a conservative low-frequency cutoff of $10\,\mathrm{Hz}$, the observations span $\sim 1-5\,\mathrm{s}$, ending with the extremely relativistic merger. 
These signals will be extremely loud, with ground-based SNRs in the range from several hundreds to thousands (see Tables~\ref{tab:supp_parameters_heavy} and \ref{tab:supp_parameters_light}
in the supplement).

We model the signals above $10\,\mathrm{Hz}$, including the effects of spin precession and higher order modes (the eccentricity is $\lesssim 10^{-5}$ and is set to zero), using \texttt{IMRPhenomXPHM}~\cite{Pratten:2020fqn,Garcia-Quiros:2020qpx,Pratten:2020ceb}.
We further assume the noise and calibration in each instrument are known at a level much better than the statistical uncertainty affecting the observations, and perform Bayesian analysis using \textsc{Bilby}~\cite{Ashton:2018jfp} (for details, see the supplement).

The large SNRs in the ground-based band allow for exquisite measurements of BH masses and spins.
In the 3G detectors, the individual masses are measured with fractional errors of $\sim 0.1\%$ and the spin magnitudes and orientations at the $1\%$ level; see Fig.~\ref{fig:intrinsic} and the supplement.  
(In the HLV+ network, errors are approximately an order of magnitude larger due to the lower SNR.)
\begin{figure}[tbp]
	\centering
	\includegraphics[width=0.95\columnwidth]{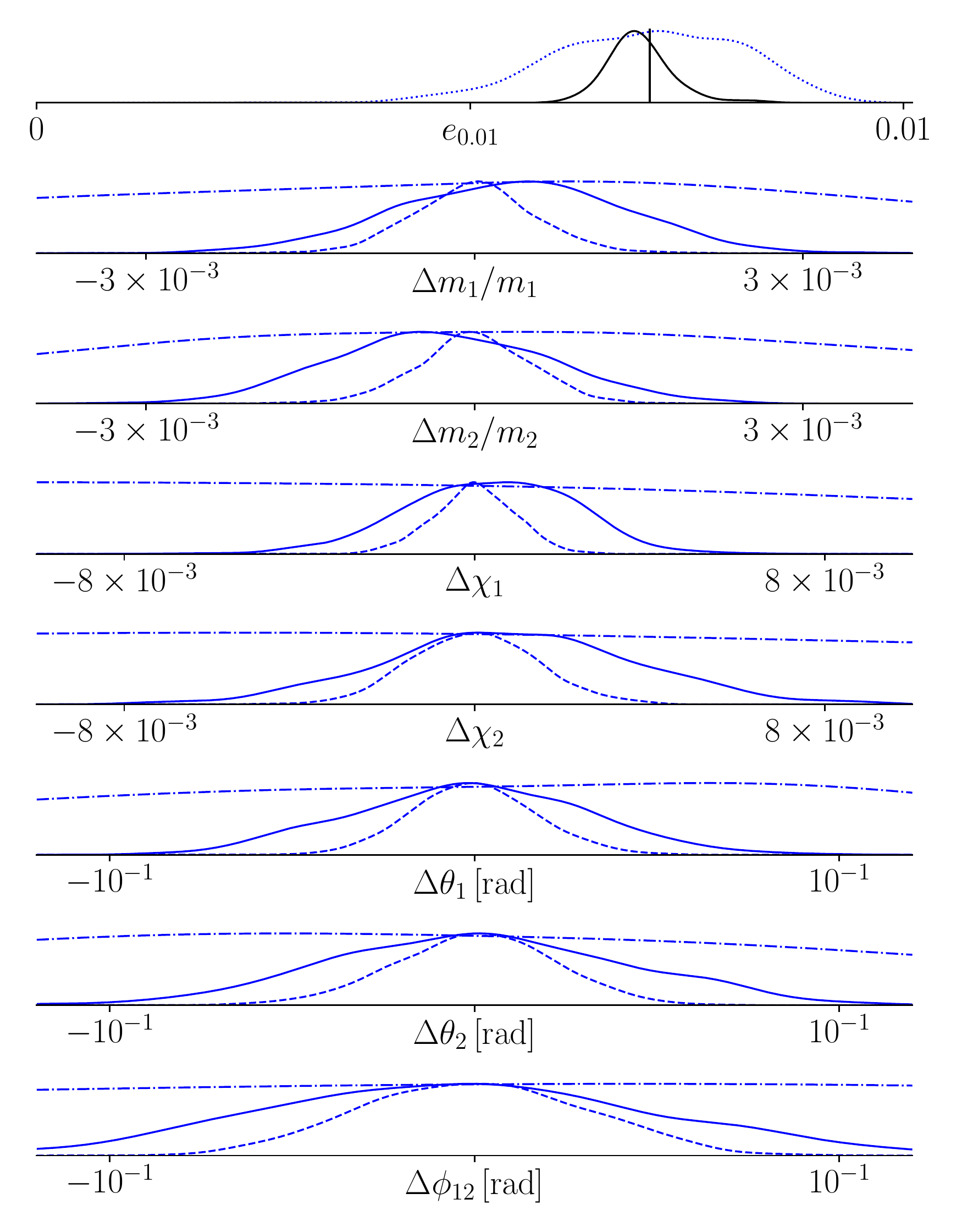}
	\caption{ \label{fig:intrinsic}
	    One-dimensional marginalised posteriors on the intrinsic parameters for the ``light'',  high-eccentricity source: 
		$10\,\mathrm{mHz}$ eccentricity $\ecc$, 
		detector frame component masses $m_{1(2)}$, 
		dimensionless spin magnitudes $\chi_{1(2)}$,
		spin tilt angles $\theta_{1(2)}$, 
		and the angle between the in-plane spins $\phi_{12}$.
		Dot/dash/solid/dot-dash lines indicate LISA/CE/ET/HLV+ posteriors. 
		LISA measures eccentricity extremely well from the early inspiral (the black curve shows the improvement with the multiband measurement of $t_m$) and the 3G instruments measure the other intrinsic parameters with exquisite accuracy.
	    The injected eccentricity is indicated; for all other parameters, $\Delta$ denotes a shift from the injected value. 
	    Similar results for the ``heavy'' source are shown in Fig.~\ref{fig:supp_Mc_tm_e0}.}
\end{figure}

\noindent{\bf \em Localisation~--~}
Multiband GW observations are expected to be restricted to a small number of high-mass, low-redshift ($z\lesssim 0.1$) sources.
The advanced warning provided by LISA makes these highly promising targets for multimessenger follow-up across the electromagnetic (EM) spectrum, and hence localisation is crucial.

The LISA observations, ending several days before merger, predict the time of the merger to within a few hours and the sky position to within a 90\% area of $\approx \SI{4}{\deg\squared}$ ($ \SI{10}{\deg\squared}$) for the light (heavy) source; see Fig.~\ref{fig:extrinsic}.
LISA also measures the distance to these sources with fractional errors of $\sim 30\%$.
Assuming the LISA data processing can be performed sufficiently quickly, there is time for telescopes to be positioned before the merger.

With ground-based observations, the localisation improves further. 
This improvement depends on the number and locations of the ground-based instruments; we show results for a single triangular-shaped ET detector and the \HLV network (we do not consider CE, as a single L-shaped instrument operating alone is not expected to determine the sky position accurately).

ET alone localises these extremely loud sources to a small area; however, the posteriors contain several widely separated modes (see Fig.~\ref{fig:extrinsic}) correlated with the GW polarisation.
Only one of these modes is consistent with the LISA measurements, and so the combined multiband (LISA-ET) network can localise the sources to within $4\,\mathrm{deg}^2$ ($0.7\,\mathrm{deg}^2$). ET can also measure the luminosity distances to fractional errors of 0.6\% (2\%) thereby localising the sources to a 3D 50\% volumes of $3\,\mathrm{Mpc}^3$ ($50\,\mathrm{Mpc}^3$).
For comparison, the \HLV network can potentially make better measurements of the sky position because of the multiple, separated instruments (see Fig.~\ref{fig:extrinsic} insets) but less accurate distance measurements due to the lower SNRs; the 3D volumes are $1\,\mathrm{Mpc}^3$ ($700\,\mathrm{Mpc}^3$).

These small localisation volumes mean that it may be possible to identify a likely host galaxy from GW observations alone, even without a counterpart. 
Integrating the galaxy luminosity function~\citep{Faber2007} over the range $0.001-10 L_*$ gives a galaxy number density of $\approx 1\,\mathrm{Mpc}^{-3}$. 
Restrict further to likely hosts with $L\gtrsim L_*$, this density reduces to $\sim 0.2\,\mathrm{Mpc}^{-3}$, giving in $\sim 1-100$ likely host galaxies in these localisation volumes.

\noindent{\bf \em Multimessenger astronomy~--~}
The form or existence of any electromagnetic counterpart to BBH mergers is currently unknown. 
If a host is identified even in the absence of a counterpart, this will yield the redshift allowing for a measurement of the Hubble constant with a statistical error $\lesssim 1\,\mathrm{km}\,\mathrm{s}^{-1}\,\mathrm{Mpc}^{-1}$ for LISA-3G multiband observations. 
In addition, identifying the host will also enable detailed measurements of the environmental metallicity and star-formation history, providing additional information on the formation channel of the binary. 
If the host contains an AGN, this would suggest accretion disks may play a role in the formation, since the number density of AGN is low at $\sim 10^{-5}\,\mathrm{Mpc}^{-3}$ \citep{Flesch2021}. 

Unique to multiband events is the few days advance warning of a merger with a localization of $\sim 1\,\mathrm{deg}^2$. 
This is well-matched to several wide-field instruments across the EM spectrum. 
On the shortest timescales, the Square Kilometre Array (SKA; field-of-view (FOV) $\sim 1\,\mathrm{deg}^2$) would place sensitive constraints on any associated fast radio burst (FRB), e.g. could detect a faint FRB -- like the 2020 outburst of SGR 1935+2154 \citep{Bochenek2020,CHIME2020} -- at $z=0.03$. 
Similar gamma-ray burst (GRB) tests could be done with missions including SVOM \citep{Cordier2015}, THESEUS \citep{Amati2018}, Gamow \citep{2021SPIE11821E..09W} and the Einstein Probe \citep{Yuan2015}, characterised by $\sim $ steradians FOV. Pointed X-ray instruments will play an important role in constraining any shock breakout transient, e.g. from a dense circumstellar medium. 
ATHENA's FOV of $\sim 0.5\,\mathrm{deg}^2$ will cover most of the localization area \citep{Rau2013,McGee2020}, and the planned ULTRASAT \citep{Sagiv2014} would provide sensitive, wide-field UV coverage.

Optical or infrared transients, if present, are expected to arise on longer timescales (from days to weeks/months), either as a result of an off-axis GRB afterglow or a thermal/diffusive transient, such as a merger within an AGN disk \citep{McKernan2019,Graham:2020gwr}.
The Rubin Observatory would be an ideal instrument for repeat monitoring to search for optical emission on a wide range of timescales, with a FOV that fully covers the localisation area and benefiting from pre-existing templates at full LSST 10-year depth to perform image subtraction for faint transient detection. A high-confidence host galaxy association enables monitoring with sensitive and high-resolution (but narrow-field) instruments, such as JWST (which may last until the 2040s after a successful launch) or ELT. 
Repeat visits could search for very faint EM emission in the optical or infrared in the weeks following merger -- a 1-hour exposure with JWST could probe 100 times deeper than a single visit with Rubin.

\begin{figure}[tbp]
	\centering
	\includegraphics[width=\columnwidth, trim={1cm 1.5cm 0 1.5cm}]{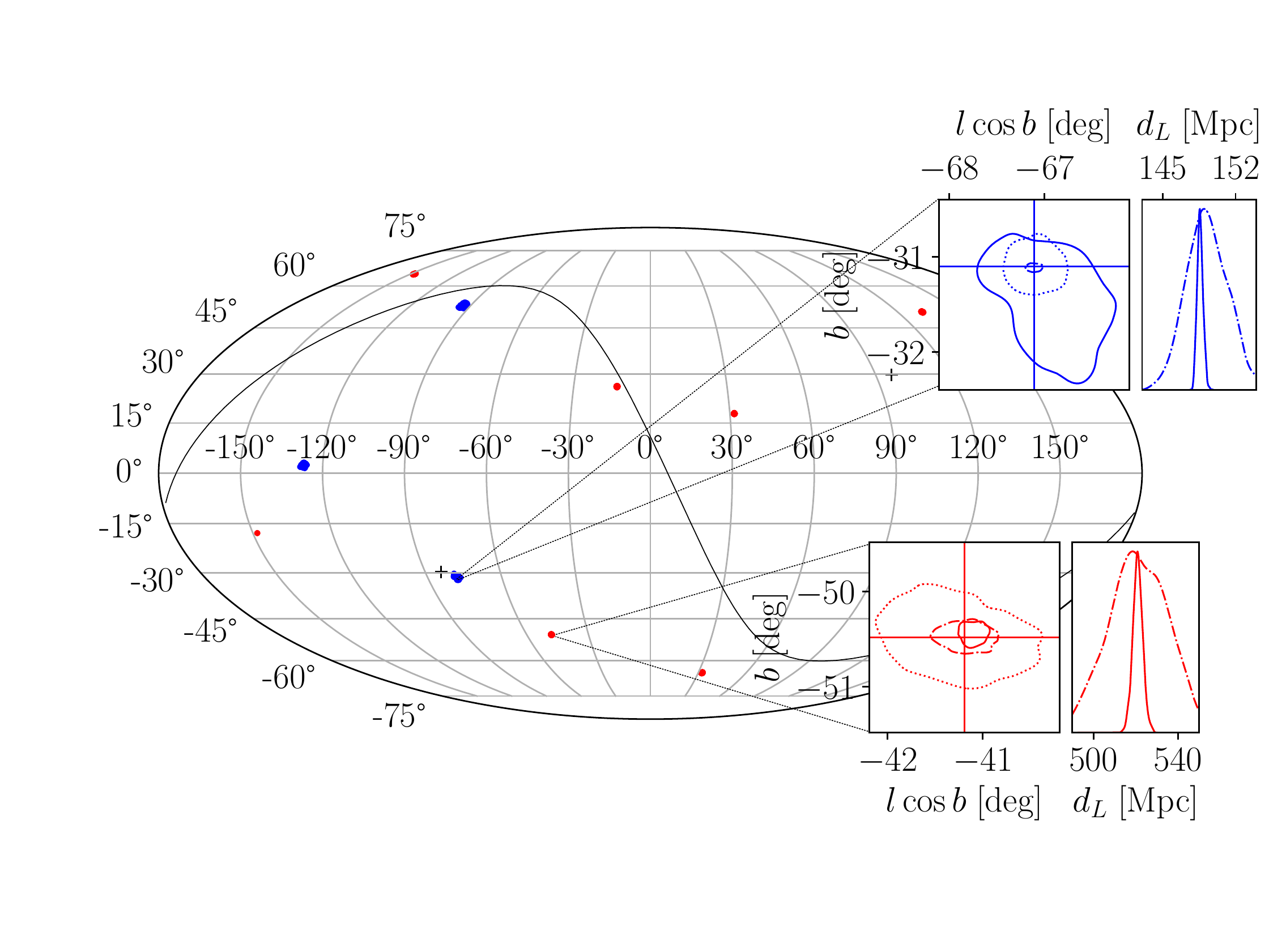}
	\caption{ \label{fig:extrinsic}
		\emph{Main panel}: Mollweide projection of the sky in galactic coordinates (ecliptic plane and poles are indicated by black line and crosses, respectively).
		Blue/red plots shows results for the light/heavy sources.
		The ET sky maps are multimodal with a small number of small, well-separated regions.
		\emph{Left insets}: Zoom-in on the true location, showing the 90\% contours of LISA/ET/\HLV measurements in dotted/solid/dot-dashed curves.
		\emph{Right insets}: Posteriors on the luminosity distance $d_L$.
	}
\end{figure}
%

\noindent{\bf \em Astrophysical implications~--~}
Only a few SmBBHs are expected to be observed first by LISA and then ground-based instruments, but the rate is uncertain and depends on the binary merger rate, the mass spectrum and the SNR detection threshold required in LISA~\cite{2019PhRvD..99j3004G, 2019MNRAS.488L..94M, 2022arXiv220102766S}.
Searches of LISA archival data targeted on systems detected with a ground-based network will further increase this number~\cite{Ewing:2020brd,2019PhRvD..99j3004G}. 

While the BH masses are highly degenerate among the many proposed formation pathways of SmBBHs, spins and eccentricities are believed to be better tracers.
SmBBH eccentricities can be excited by dynamical encounters and are thus unique observables to probe black-hole multi-body dynamics. 
Circularization highly suppresses the eccentricity in the ground-based band, but observing SmBBHs in the LISA band provides a new, powerful leverage \cite{2016ApJ...830L..18B,2017MNRAS.465.4375N}. 
At $\sim10\,\mathrm{mHz}$, binaries formed in galactic fields (dense clusters) are expected to have eccentricities of $\sim 10^{-4}\! -\! 10^{-3}$ ($10^{-3}\!-\!10^{-1}$) \cite{2017MNRAS.465.4375N}.
Kozai-Lidov oscillations induced by a nearby supermassive black-hole can increase the SmBBH's eccentricity to $\ecc \lesssim 1$~\cite{2012ApJ...757...27A}.
SmBBH spin directions also encode information on the evolutionary mechanism~\cite{2013PhRvD..87j4028G,2016ApJ...832L...2R,2017MNRAS.471.2801S,2018PhRvD..98h4036G}. 
Dynamical interactions should randomize the spin orientations. 
On the other hand, binaries formed in isolation will be subject to small misalignment $\lesssim 20^\circ$ \cite{2018PhRvD..98h4036G} at merger, induced by e.g. supernova kicks \cite{2000ApJ...541..319K} and tidal interactions~\cite{1981A&A....99..126H}. 
Spin misalignments in the ground-based and LISA bands are strongly correlated~\cite{2015PhRvD..92f4016G}.

Our results show that LISA can measure $\ecc$ with absolute errors of $\sim 10^{-4} - 10^{-3}$ while 3G detector will provide estimates of the spin tilts down to $\sim 1\mathrm{deg}$. 
The resulting astrophysical inference from either of the measurements alone will already be incredibly powerful, but a consistent determination of the formation channel from two observables (spins and eccentricity) and two bands (ground and space) will open a new era of precision black-hole binary astrophysics. 
Identification of the galactic host would allow for precise determination of the metallicity of the environment where the source evolved. This will place a further constraint in pinning down the details of the formation mechanism (e.g. \cite{2018MNRAS.480.2011G}), possibly breaking the mass degeneracy that affects the proposed pathways.

\noindent{\bf \em Tests of Fundamental Physics~--~}
The prospect of multiband GW observations spanning several orders of magnitude in frequency would seem to offer a unique opportunity for testing general relativity (GR) across a wide range of scales with a single source~\cite{2021PhRvD.103d4024P}.
Indeed, several papers have suggested that these multiband sources will be ideal for testing aspects of fundamental physics such as parametrized post-Einsteinian deviations~\cite{2009PhRvD..80l2003Y}, inspiral-merger-ringdown consistency tests \cite{2016PhRvL.117e1102V, 2019PhRvD.100f4024G, 2020PhRvD.101d4047C}, measurements of post-Newtonian coefficients \cite{2020PhRvL.125t1101G, 2021PhRvD.103b4036D}, and the existence of dipole GW radiation \cite{2016PhRvL.116x1104B, 2017PhRvD..96h4039C, 2020MNRAS.496..182L}.

However, our results suggest these sources are not as good for testing GR as had been initially hoped, for several related reasons.
Firstly, most previous studies have relied on Fisher matrices, which are not well-suited to these sources, particularly in the LISA band \cite{Toubiana:2020vtf}.
Secondly, the low SNR of the LISA observations, compared to the ground-based instruments, means some multiband tests are limited by LISA and are no more constraining than what is achievable with 3G ground-based detectors alone, or with high-SNR LISA observations of supermassive BBHs.
This was also suggested in Ref.~\cite{Toubiana:2020vtf}.

Consider the possibility of dipole emission (which can occur in modified gravity theories, or if the BHs are electrically charged). 
It dominates at low frequencies, accelerating the inspiral and reducing the time to merger.
It has been suggested that multiband sources, where the merger time can be predicted from the low-frequency signal, should be ideal for constraining dipole emission.
Ref.~\cite{2017PhRvD..96h4039C}, using a Fisher matrix analysis, found that LISA can predict the time of merger for a GW150914-like source to within $\pm 10\,\mathrm{s}$ and place a constraint on the size of a dipole contribution to the flux that is an order of magnitude better than is possible with other GW sources (although still worse than the double pulsar constraint~\cite{2006Sci...314...97K} for relevant theories). 
However, as has been shown here with a full Bayesian analysis, the LISA determination of the merger time is significantly worse at around $\pm 3\,\mathrm{hours}$ (see Fig.~\ref{fig:Mc_e0_tm}). 
As the dipole constraint degrades linearly with the time of merger uncertainty, the constraints on the dipole could be less constraining than has been previously suggested.
This should be verified with a Bayesian analysis incorporating modifications of GR.
We find that for our light source, the dimensionless dipole coupling $B$ \cite{2016PhRvL.116x1104B} can be constrained to $\lesssim 10^{-7}$;
this corresponds to a constraint on the BH charges of $Q\lesssim 2 \times 10^{-4} Q_{\rm max}$, where $Q_{\rm max}$ is the extremal BH charge.

\noindent{\bf \em Challenges to be met~--~}
These multi-band observations pose challenges on data processing, waveform modelling and instrument calibration that require work to start immediately. 

The early warning that enables telescope pointing ahead of the merger requires the LISA data processing to be done in near real time and completed within $\sim$ a day from the time a SmBBH candidate signal exits the LISA band. 
We currently do not have a search strategy for SmBBHs, although solutions are being developed as part of the LISA Data Challenges~\cite{ldc}. 
Accurate measurements of the coalescence time and sky position with the fully Bayesian techniques used here are highly computationally expensive (see also \cite{balrog-ldc1}). This combination of factors drives demanding requirements on LISA data processing and data transfer. Search strategies also need to be developed in the context of LISA archival searches for signals detected in ground-based data sets. 

The exquisitely precise characterisation of the properties of these sources requires waveforms for eccentric and precessing systems that are phase coherent over $\sim 10^6$ binary orbits and $\sim 10^3$ precession and periapsis precession cycles~\cite{2019PhRvD..99f4056M}. Moreover, the extreme SNRs of $\sim 10^2-10^3$ in the ground-based observations, where one should also account for the possible presence of small residual eccentricities, places requirements on waveform accuracy that go well beyond what is currently available.

Equally, one needs to ensure that other modelling components of the GW signal as measured at the detector output introduces systematic errors below the statistical ones. Specifically, in our analysis, we have not considered uncertainties coming from instrument calibration errors or unknown noise levels. We have \textit{de-facto} assumed calibration is exact. Calibration errors will need to be  smaller than $0.1\%$ in amplitude and  $\lesssim 0.1\,\mathrm{deg}$ in phase. 

Finally, we want to point out that there is a blind (in frequency) spot in observing these sources at $\sim 0.1 - 1\,\mathrm{Hz}$. Several space-based mission concepts that can cover this frequency band are currently under study~\cite{2016CQGra..33c5010L, 2018arXiv180709495R, 2021PTEP.2021eA105K}. Although the impact of the operation of these instruments concurrently with LISA and ground-based instruments is beyond the scope of this paper, unquestionably they would provide additional information.

\noindent{\bf \em Acknowledgments~--~}
%
We thank all the developers of the \textsc{Balrog} codesuite, including those who are not authors here: Janna Goldstein, Elinore Roebber, Sebastian Gaebel, Siyuan Chen, Christopher Berry.
Computational resources used for this work were provided by the University of Birmingham's BlueBEAR High Performance Computing facility and by Supercomputing Wales, funded by STFC grants ST/I006285/1 and ST/V001167/1 supporting UK involvement in the operation of Advanced LIGO.
A.K., C.J.M. and A.V. acknowledge the support of the UK Space Agency, Grant No. ST/V002813/1.
A.V. acknowledges the support of the Royal Society and Wolfson Foundation.
R.B. is supported by the Italian Space Agency Grant ``Phase A LISA mission activities'', Agreement No.~2017-29-H.0, CUP~F62F17000290005.
D.G. is supported by European Union's H2020 ERC Starting Grant No.~945155--GWmining, Cariplo Foundation Grant No.~2021-0555, and Leverhulme Trust Grant No.~RPG-2019-350.
P. S., G.P. and A.V. acknowledge support from STFC grant ST/V005677/1. 
M.N. is supported by the European Research Council (ERC) under the European Union’s Horizon 2020 research and innovation programme (grant agreement No.~948381) and by a Fellowship from the Alan Turing Institute.
Part of this research was performed while G.P., P.S. and L.T. were visiting the Institute for Pure and Applied Mathematics (IPAM), supported by the NSF, Grant No. DMS-1925919.

\bibliographystyle{apsrev4-2}
\bibliography{refs}

\clearpage
\newpage

\setcounter{equation}{0}
\setcounter{figure}{0}
\setcounter{table}{0}
\renewcommand{\theequation}{S\arabic{equation}}
\renewcommand{\thefigure}{S\arabic{figure}}
\renewcommand{\thetable}{S\arabic{table}}

\setcounter{page}{1}


\title{Supplemental material: ``The last three years: multiband
gravitational-wave observations of stellar-mass binary black holes''}
\maketitle

\onecolumngrid
\section{System parameters}\label{supp:par}

Here, we provide detailed values of the source parameters selected for our inference. They are summarised in Tables~\ref{tab:supp_parameters_heavy} and~\ref{tab:supp_parameters_light}.

In the LISA band, we perform four distinct injections of gravitational-wave signals from binary systems. 
Their parameters are consistent with those of GW150914 (``\emph{light}'') and GW190521 (``\emph{heavy}''), with the exception of the luminosity distance that we chose such that the binaries are near the (currently understood) detection threshold SNR in LISA. 
In the detector frame, the \emph{light} (\emph{heavy}) binary has a primary component mass $m_1=\SI{36}{\solarmass}(\SI{85}{\solarmass})$ and a secondary component of $m_2=\SI{29}{\solarmass} (\SI{65}{\solarmass})$.
Correspondingly, their chirp mass is $\mathcal{M}_c=\SI{28.1}{\solarmass} (\SI{64.6}{\solarmass})$.
The luminosity distance is set to $d_L=\SI{149.1}{\mega\parsec}~(\SI{521.10}{\mega\parsec})$, in order to give an ${\rm SNR=11.9~(15.3)}$ for LISA over a mission duration of $\SI{4}{\year}$.

Ecliptic longitude and latitude are set to $l=\SI{5.01}{\radian}$ and $b=\SI{-1.49}{\radian}$ ($l=\SI{5.55}{\radian}$ and $b=\SI{-0.83}{\radian}$), respectively, and the inclination to $\SI{2.28}{\radian}$ ($\SI{0.68}{\radian}$).
The binary is set to an initial orbital frequency such that it reaches merger $\SI{1e+8}{\second}$ (approximately 3 years) after the start of the LISA observation.
For convenience, times are stated shifted such that the merger occurs at $t_m=0$.

The signals are injected with a small but non-zero eccentricity. We perform injections with two different eccentricities (``\textit{high}'' and ``\textit{low}'') for each binary: $\num{7.7e-3}$ and $\num{1e-4}$ for the \textit{light} binary and $\num{3.16e-2}$ and $\num{1e-4}$ for the \textit{heavy} binary.
This spread of values was chosen to illustrate our ability to confidently distinguish small values from zero --or not-- with LISA observations.

Conserved quantities are used consistently to generate and inject signals for the ground-based detector networks.
Non-conserved parameters (spin and orbital angular momentum magnitudes and angles, and eccentricities) are defined at $\SI{0.01}{\hertz}$ for LISA signals. 
Dimensionless spin magnitudes amount to 
$\chi_1=c |\vec{S_1}|/Gm_1^2=0.1329$ 
and
$\chi_2=c |\vec{S_2}|/Gm_2^2=0.1384$ 
($\chi_1=0.7616$ 
and 
$\chi_2=0.8545$).
Spin tilts relative to the respective binary orbital angular momentum are 
$\theta_1=\arccos(\hat{S}_1\cdot\hat{L})=\SI{\LISALightLoEccSpinOneTilt}{\radian}$ 
and 
$\theta_2=\arccos(\hat{S}_2\cdot\hat{L})=\SI{\LISALightLoEccSpinTwoTilt}{\radian}$ 
($\theta_1= \SI{\LISAHeavyLoEccSpinOneTilt}{\radian}$ 
and 
$\theta_2= \SI{\LISAHeavyLoEccSpinTwoTilt}{\radian}$), respectively.
The resulting angle between the line-of-sight unit vector and the total angular momentum is set to 
$\theta_\text{JN}= \SI{\LISALightLoEccThetaJN}{\radian}$ 
($\theta_\text{JN}= \SI{\LISAHeavyLoEccThetaJN}{\radian}$) 
and
$\theta_\text{JN}= \SI{\LISALightHiEccThetaJN}{\radian}$ 
($\theta_{\text{JN}}= \SI{\LISAHeavyHiEccThetaJN}{\radian}$) 
for the low-eccentricity and high-eccentricity binaries, respectively.
Analogously, the azimuthal angle of $\hat{L}$ with respect to $\hat{J}$ is set to 
$\phi_{\text{JL}}= \SI{\LISALightLoEccPhiJL}{\radian}$ 
($\phi_{\text{JL}}= \SI{\LISAHeavyLoEccPhiJL}{\radian}$)
and
$\phi_{\text{JL}}= \SI{\LISALightHiEccPhiJL}{\radian}$ 
($\phi_{\text{JL}}= \SI{\LISAHeavyHiEccPhiJL}{\radian}$).
These parameters are then evolved to $\SI{10}{\hertz}$ for injection in the ground-based detector signals
\cite{2010PhRvD..81l4001K, Klein:2021jtd}. 
Dimensionless spin magnitudes for ground-based detector signals are set to $\chi_1=0.1329$ and $\chi_2=0.1383$ ($\chi_1=0.7615$ and $\chi_2=0.8544$). Spin tilts with respect to the binary orbital angular momentum are $\theta_1= \SI{1.53}{\radian}$ and $\theta_2= \SI{1.50}{\radian}$ ($\theta_1= \SI{0.83}{\radian}$ and $\theta_2= \SI{1.85}{\radian}$).
The resulting angle between the line-of-sight unit vector and the total angular momentum is set to $\theta_{\text{JN}}= \SI{2.31}{\radian}$ ($\theta_{\text{JN}}= \SI{0.88}{\radian}$), while the azimuthal angle of $\hat{L}$ with respect to $\hat{J}$ is set to $\phi_{\text{JL}}= \SI{0.77}{\radian}$ ($\phi_{\text{JL}}= \SI{2.03}{\radian}$).
The eccentricity is set to zero in the ground-based band.

Except for the luminosity distance, these parameter choices are consistent with the posteriors of GW150914~\cite{gwoscGW150914} and GW190521~\cite{gwoscGWTC-2.1}, respectively.
 
\begin{table}
\begin{center}
\begin{tabular*}{0.851\textwidth}{|c|cc|ccccc|}
    \hline
    \hline
    \multicolumn{8}{|c|}{ }\\
    \multicolumn{8}{|c|}{Heavy binary}\\
    \multicolumn{8}{|c|}{ }\\
    \hline
    & & & & & & & \\
    &\multicolumn{2}{c|}{Injected value}& \multicolumn{5}{c|}{Posterior} \\ 
    & & & & & & & \\ \cline{2-8}
    & & & & & & & \\
    & \multicolumn{2}{c|}{\phantom{00}10mHz\phantom{00.000}10Hz\phantom{0}} & \multicolumn{2}{c}{LISA} & ET & CE & HLV+ \\ 
    & & & & & & & \\
    & Eccentricity & & High ecc. & Low ecc. & & & \\
    & High\phantom{ }$\mid$\phantom{ }Low& & & & & & \\
    \hline
    & & & & & & & \\
    $\rho_{\text{opt}}$ & & & $\HeavyHiEccSNRRecovered$ & $\HeavyLoEccSNRRecovered$ & $\ETDHeavyOptSNRRecovered$&$\CEHeavyOptSNRRecovered$&$\HLVHeavyOptSNRRecovered$\\
    & & & & & & & \\
    $\mathcal{M}_{\text{c}}$ $\left[M_{\odot}\right]$&\multicolumn{2}{c|}{$\HeavyChirpMassInjected$}&$\HeavyHiChirpMassRecovered$&$\HeavyLoChirpMassRecovered$&$\ETDHeavyChirpMassRecovered$&$\CEHeavyChirpMassRecovered$&$\HLVHeavyChirpMassRecovered$\\
    & & & & & & & \\
    $M_{\texttt{tot}}$ $\left[M_{\odot}\right]$& \multicolumn{2}{c|}{$\HeavyTotalMassInjected$}&$\HeavyHiTotalMassRecovered$&$\HeavyLoTotalMassRecovered$&$\ETDHeavyTotalMassRecovered$&$\CEHeavyTotalMassRecovered$&$\HLVHeavyTotalMassRecovered$\\
    & & & & & & & \\
    $q$&\multicolumn{2}{c|}{$\HeavyMassRatioInjected$}&$\HeavyHiMassRatioRecovered$&$\HeavyLoMassRatioRecovered$&$\ETDHeavyMassRatioRecovered$&$\CEHeavyMassRatioRecovered$&$\HLVHeavyMassRatioRecovered$\\
    & & & & & & & \\
    $m_{1}$ $\left[M_{\odot}\right]$&\multicolumn{2}{c|}{$\HeavyMassOneInjected$}&$\HeavyHiMassOneRecovered$&$\HeavyLoMassOneRecovered$&$\ETDHeavyMassOneRecovered$&$\CEHeavyMassOneRecovered$&$\HLVHeavyMassOneRecovered$\\
    & & & & & & & \\
    $m_{2}$ $\left[M_{\odot}\right]$&\multicolumn{2}{c|}{$\HeavyMassTwoInjected$}&$\HeavyHiMassTwoRecovered$&$\HeavyLoMassTwoRecovered$&$\ETDHeavyMassTwoRecovered$&$\CEHeavyMassTwoRecovered$&$\HLVHeavyMassTwoRecovered$\\
    & & & & & & & \\
    $\chi_1$&\multicolumn{2}{c|}{$\HeavyAOneInjected$}&$\HeavyHiAOneRecovered$&$\HeavyLoAOneRecovered$&$\ETDHeavyAOneRecovered$&$\CEHeavyAOneRecovered$&$\HLVHeavyAOneRecovered$\\
    & & & & & & & \\
    $\chi_2$&\multicolumn{2}{c|}{$\HeavyATwoInjected$}&$\HeavyHiATwoRecovered$&$\HeavyLoATwoRecovered$&$\ETDHeavyATwoRecovered$&$\CEHeavyATwoRecovered$&$\HLVHeavyATwoRecovered$\\
    & & & & & & & \\
    $\iota$ $\left[\text{rad}\right]$& $\LISAHeavyHiEccIota$ &$\HeavyIotaInjected$& $\HeavyHiIotaRecovered$ & $\HeavyLoIotaRecovered$ &\mm&\mm&$\HLVHeavyIotaRecovered$\\
    & & & & & & & \\
    $\theta_1$ $\left[\text{rad}\right]$& $\LISAHeavyLoEccSpinOneTilt$ &$\HeavyTiltOneInjected$&$\HeavyHiTiltOneRecovered$&$\HeavyLoTiltOneRecovered$&$\ETDHeavyTiltOneRecovered$&$\CEHeavyTiltOneRecovered$&$\HLVHeavyTiltOneRecovered$\\
    & & & & & & & \\
    $\theta_2$ $\left[\text{rad}\right]$& $\LISAHeavyLoEccSpinTwoTilt$ &$\HeavyTiltTwoInjected$&$\HeavyHiTiltTwoRecovered$&$\HeavyLoTiltTwoRecovered$&$\ETDHeavyTiltTwoRecovered$&$\CEHeavyTiltTwoRecovered$&$\HLVHeavyTiltTwoRecovered$\\
    & & & & & & & \\
    $\phi_{12}$ $\left[\text{rad}\right]$& $\LISAHeavyLoEccPhiOneTwo$ &$\HeavyPhiOneTwoInjected$&$\HeavyHiPhiOneTwoRecovered$&$\HeavyLoPhiOneTwoRecovered$&$\ETDHeavyPhiOneTwoRecovered$&$\CEHeavyPhiOneTwoRecovered$&$\HLVHeavyPhiOneTwoRecovered$\\
    & & & & & & & \\
    $\phi_{\text{JL}}$ $\left[\text{rad}\right]$& $\LISAHeavyHiEccPhiJL \mid \LISAHeavyLoEccPhiJL$ &$\HeavyPhiJLInjected$& $\HeavyHiPhiJLRecovered$
 & $\HeavyLoPhiJLRecovered$  &\mm&\mm&$\HLVHeavyPhiJLRecovered$\\
    & & & & & & & \\
    $\theta_{\text{JN}}$ $\left[\text{rad}\right]$&$\LISAHeavyHiEccThetaJN\mid \LISAHeavyLoEccThetaJN$ &$\HeavyThetaJNInjected$& $\HeavyHiThetaJNRecovered$&$\HeavyLoThetaJNRecovered$ &\mm&\mm&$\HLVHeavyThetaJNRecovered$\\
    & & & & & & & \\
    $\chi_{\text{p}}$& $\LISAHeavyHiEccChip$ &$\HeavyChiPInjected$&$\HeavyHiChiPRecovered$&$\HeavyLoChiPRecovered$&$\ETDHeavyChiPRecovered$&$\CEHeavyChiPRecovered$&$\HLVHeavyChiPRecovered$\\
    & & & & & & & \\
    $\chi_{\text{eff}}$&$\LISAHeavyHiEccChieff$&$\HeavyChiEffInjected$&$\HeavyHiChiEffRecovered$&$\HeavyLoChiEffRecovered$&$\ETDHeavyChiEffRecovered$&$\CEHeavyChiEffRecovered$&$\HLVHeavyChiEffRecovered$\\
    & & & & & & & \\
    $e_{0.01}$& $\HeavyHiEccInjected \mid \HeavyLoEccInjected$& N/A & $\HeavyHiEccRecovered$ & $\HeavyLoEccRecovered$ & N/A & N/A & N/A \\
    & & & & & & & \\
    $d_L$ $\left[\text{Mpc}\right]$&\multicolumn{2}{c|}{$\HeavyLuminosityDistanceInjected$}&$\HeavyHiLuminosityDistanceRecovered$&$\HeavyLoLuminosityDistanceRecovered$&$\ETDHeavyLuminosityDistanceRecovered$&\mm&$\HLVHeavyLuminosityDistanceRecovered$\\
    & & & & & & & \\
    $z$&\multicolumn{2}{c|}{$\HeavyRedshiftInjected$}&$\HeavyHiRedshiftRecovered$&$\HeavyLoRedshiftRecovered$&$\ETDHeavyRedshiftRecovered$&\mm&$\HLVHeavyRedshiftRecovered$\\
    & & & & & & & \\
    $\alpha$ (RA)$\left[\text{rad}\right]$&\multicolumn{2}{c|}{$\HeavyRAInjected$}&$\HeavyHiEccRARecovered$&$\HeavyLoEccRARecovered$&\mm&\mm&$\HLVHeavyRARecovered$\\
    & & & & & & & \\
    $\delta$ (DEC)$\left[\text{rad}\right]$&\multicolumn{2}{c|}{$\HeavyDECInjected$}&$\HeavyHiEccDECRecovered$&$\HeavyLoEccDECRecovered$&\mm&\mm&$\HLVHeavyDECRecovered$\\
    & & & & & & & \\
    $\Delta t_m$ $\left[\text{s}\right]$& \multicolumn{2}{c|}{0} &$\HeavyHiEccCoalTimeRecovered$&$\HeavyLoEccCoalTimeRecovered$&\mm&\mm&$\HLVHeavyCoalTimeRecovered$\\
    & & & & & & & \\  
    $\Delta \Omega$ $\left[\text{sq. deg.}\right]$&\multicolumn{2}{c|}{N/A}&$\LISAHeavyHiEccOmega$&$\LISAHeavyLoEccOmega$&$\ETDHeavyOmega$&N/A&$\HLVHeavyOmega$\\
    & & & & & & & \\
   \hline
   \hline
\end{tabular*}
\end{center}
 \caption[width=\textwidth]{
   Injected and recovered source parameters for the heavy binary.
   For most parameters, we choose to quote the median posterior values with the 90\% confidence intervals in either of the equivalent notations $0.1234^{+5}_{-5}=0.1234(5)$, as appropriate.
   For some parameters that can't be confidently measured to be away from a prior boundary, we instead choose to quote 90\% limits.
   No values are given for parameters where posteriors are extremely multi-modal (MM) and where it doesn't make sense to quote a single range; we refer the interested reader to the plots instead.
   All spin-related quantities are quoted at a two reference frequencies, $10\,\mathrm{Hz}$ and $10\,\mathrm{mHz}$, corresponding to the LISA and ground-based bands.
   Eccentricities are quoted at a reference frequency of $10\,\mathrm{mHz}$.
   \label{tab:supp_parameters_heavy}}
\end{table}

\begin{table}
\begin{center}
\begin{tabular*}{0.885\textwidth}{|c|cc|ccccc|}
    \hline
    \hline
    \multicolumn{8}{|c|}{ }\\
    \multicolumn{8}{|c|}{Light binary}\\
    \multicolumn{8}{|c|}{ }\\
    \hline
    & & & & & & & \\
    &\multicolumn{2}{c|}{Injected value}& \multicolumn{5}{c|}{Posterior} \\ 
    & & & & & & & \\ \cline{2-8}
    & & & & & & & \\
    & \multicolumn{2}{c|}{\phantom{00}10mHz\phantom{00.000}10Hz\phantom{0}} & \multicolumn{2}{c}{LISA} & ET & CE & HLV+ \\ 
    & & & & & & & \\
    & Eccentricity & & High ecc. & Low ecc. & & & \\
    & High\phantom{ }$\mid$\phantom{ }Low& & & & & & \\
    \hline
    & & & & & & & \\
    $\rho_{\text{opt}}$ & & & $\LightHiEccSNRRecovered$ & $\LightLoEccSNRRecovered$ & $\ETDLightOptSNRRecovered$&$\CELightOptSNRRecovered$&$\HLVLightOptSNRRecovered$\\
    & & & & & & & \\
    $\mathcal{M}_{\text{c}}$ $\left[M_{\odot}\right]$&\multicolumn{2}{c|}{$\LightChirpMassInjected$}&$\LightHiChirpMassRecovered$&$\LightLoChirpMassRecovered$&$\ETDLightChirpMassRecovered$&$\CELightChirpMassRecovered$&$\HLVLightChirpMassRecovered$\\
    & & & & & & & \\
    $M_{\texttt{tot}}$ $\left[M_{\odot}\right]$& \multicolumn{2}{c|}{$\LightTotalMassInjected$}&$\LightHiTotalMassRecovered$&$\LightLoTotalMassRecovered$&$\ETDLightTotalMassRecovered$&$\CELightTotalMassRecovered$&$\HLVLightTotalMassRecovered$\\
    & & & & & & & \\
    $q$&\multicolumn{2}{c|}{$\LightMassRatioInjected$}&$\LightHiMassRatioRecovered$&$\LightLoMassRatioRecovered$&$\ETDLightMassRatioRecovered$&$\CELightMassRatioRecovered$&$\HLVLightMassRatioRecovered$\\
    & & & & & & & \\
    $m_{1}$ $\left[M_{\odot}\right]$&\multicolumn{2}{c|}{$\LightMassOneInjected$}&$\LightHiMassOneRecovered$&$\LightLoMassOneRecovered$&$\ETDLightMassOneRecovered$&$\CELightMassOneRecovered$&$\HLVLightMassOneRecovered$\\
    & & & & & & & \\
    $m_{2}$ $\left[M_{\odot}\right]$&\multicolumn{2}{c|}{$\LightMassTwoInjected$}&$\LightHiMassTwoRecovered$&$\LightLoMassTwoRecovered$&$\ETDLightMassTwoRecovered$&$\CELightMassTwoRecovered$&$\HLVLightMassTwoRecovered$\\
    & & & & & & & \\
    $\chi_1$&\multicolumn{2}{c|}{$\LightAOneInjected$}&$\LightHiAOneRecovered$&$\LightLoAOneRecovered$&$\ETDLightAOneRecovered$&$\CELightAOneRecovered$&$\HLVLightAOneRecovered$\\
    & & & & & & & \\
    $\chi_2$&\multicolumn{2}{c|}{$\LightATwoInjected$}&$\LightHiATwoRecovered$&$\LightLoATwoRecovered$&$\ETDLightATwoRecovered$&$\CELightATwoRecovered$&$\HLVLightATwoRecovered$\\
    & & & & & & & \\
    $\iota$ $\left[\text{rad}\right]$& $\LISALightHiEccIota$ &$\LightIotaInjected$& $\LightHiIotaRecovered$& $\LightLoIotaRecovered$&$ \ETDLightIotaRecovered $& $\CELightIotaRecovered $ & $\HLVLightIotaRecovered$\\
    & & & & & & & \\
    $\theta_1$ $\left[\text{rad}\right]$&$\LISALightLoEccSpinOneTilt$ &$\LightTiltOneInjected$&$\LightHiTiltOneRecovered$&$\LightLoTiltOneRecovered$&$\ETDLightTiltOneRecovered$&$\CELightTiltOneRecovered$&$\HLVLightTiltOneRecovered$\\
    & & & & & & & \\
    $\theta_2$ $\left[\text{rad}\right]$&$\LISALightLoEccSpinTwoTilt$ &$\LightTiltTwoInjected$&$\LightHiTiltTwoRecovered$&$\LightLoTiltTwoRecovered$&$\ETDLightTiltTwoRecovered$&$\CELightTiltTwoRecovered$&$\HLVLightTiltTwoRecovered$\\
    & & & & & & & \\
    $\phi_{12}$ $\left[\text{rad}\right]$&$\LISALightLoEccPhiOneTwo$ &$\LightPhiOneTwoInjected$&$\LightHiPhiOneTwoRecovered$&$\LightLoPhiOneTwoRecovered$&$\ETDLightPhiOneTwoRecovered$&$\CELightPhiOneTwoRecovered$&$\HLVLightPhiOneTwoRecovered$\\
    & & & & & & & \\
    $\phi_{\text{JL}}$ $\left[\text{rad}\right]$& $\LISALightHiEccPhiJL \mid\LISALightLoEccPhiJL$ &$\LightPhiJLInjected$& $\LightHiPhiJLRecovered$
 & $\LightLoPhiJLRecovered$&$\ETDLightPhiJLRecovered$&$\CELightPhiJLRecovered$&\mm\\
    & & & & & & & \\
    $\theta_{\text{JN}}$ $\left[\text{rad}\right]$& $\LISALightHiEccThetaJN\mid\LISALightLoEccThetaJN$ &$\LightThetaJNInjected$& $\LightHiThetaJNRecovered$& $\LightLoThetaJNRecovered$ &$\ETDLightThetaJNRecovered$&$\CELightThetaJNRecovered$&$\HLVLightThetaJNRecovered$\\
    & & & & & & & \\
    $\chi_{\text{p}}$& $\LISALightHiEccChip$ &$\LightChiPInjected$&$\LightHiChiPRecovered$&$\LightLoChiPRecovered$&$\ETDLightChiPRecovered$&$\CELightChiPRecovered$&$\HLVLightChiPRecovered$\\
    & & & & & & & \\
    $\chi_{\text{eff}}$& $\LISALightHiEccChieff$& $\LightChiEffInjected$&$\LightHiChiEffRecovered$&$\LightLoChiEffRecovered$&$\ETDLightChiEffRecovered$&$\CELightChiEffRecovered$&$\HLVLightChiEffRecovered$\\
    & & & & & & & \\
    $e_{0.01}$& $\LightHiEccInjected \mid \LightLoEccInjected$& N/A & $\LightHiEccRecovered$ & $\LightLoEccRecovered$ & N/A & N/A & N/A \\
    & & & & & & & \\
    $d_L$ $\left[\text{Mpc}\right]$&\multicolumn{2}{c|}{$\LightLuminosityDistanceInjected$}&$\LightHiLuminosityDistanceRecovered$&$\LightLoLuminosityDistanceRecovered$&$\ETDLightLuminosityDistanceRecovered$&$\CELightLuminosityDistanceRecovered$&$\HLVLightLuminosityDistanceRecovered$\\
    & & & & & & & \\
    $z$&\multicolumn{2}{c|}{$\LightRedshiftInjected$}&$\LightHiRedshiftRecovered$&$\LightLoRedshiftRecovered$&$\ETDLightRedshiftRecovered$&$\CELightRedshiftRecovered$&$\HLVLightRedshiftRecovered$\\
    & & & & & & & \\
    $\alpha$ (RA)$\left[\text{rad}\right]$&\multicolumn{2}{c|}{$\LightRAInjected$}&$\LightHiEccRARecovered$&$\LightLoEccRARecovered$&\mm&\mm&$\HLVLightRARecovered$\\
    & & & & & & & \\
    $\delta$ (DEC)$\left[\text{rad}\right]$&\multicolumn{2}{c|}{$\LightDECInjected$}&$\LightHiEccDECRecovered$&$\LightLoEccDECRecovered$&\mm&\mm&$\HLVLightDECRecovered$\\
    & & & & & & & \\
    $\Delta t_m$ $\left[\text{s}\right]$& \multicolumn{2}{c|}{0} &$\LightHiEccCoalTimeRecovered$&$\LightLoEccCoalTimeRecovered$&$\ETDLightCoalTimeRecovered$&\mm&$\HLVLightCoalTimeRecovered$\\
    & & & & & & & \\  
    $\Delta \Omega$ $\left[\text{sq. deg.}\right]$&\multicolumn{2}{c|}{N/A}&$\LISALightHiEccOmega$&$\LISALightLoEccOmega$&$\ETDLightOmega$&N/A&$\HLVLightOmega$\\
    & & & & & & & \\
   \hline
   \hline
\end{tabular*}
 \caption[width=\textwidth]{
   Injected and recovered source parameters for the light binary.
   For most parameters, we choose to quote the median posterior values with the 90\% confidence intervals in either of the equivalent notations $0.1234^{+5}_{-5}=0.1234(5)$, as appropriate.
   For some parameters that can't be confidently measured to be away from a prior boundary, we instead choose to quote 90\% limits.
   No values are given for parameters where posteriors are extremely multi-modal (MM) and where it doesn't make sense to quote a single range; we refer the interested reader to the plots instead.
   All spin-related quantities are quoted at a two reference frequencies, $10\,\mathrm{Hz}$ and $10\,\mathrm{mHz}$, corresponding to the LISA and ground-based bands.
   Eccentricities are quoted at a reference frequency of $10\,\mathrm{mHz}$.
   \label{tab:supp_parameters_light}}
\end{center}
\end{table}

\section{LISA data analysis}\label{supp:LISA}

\textbf{Detector} --
We consider an equal-arms constellation with arm-length $L=$\SI{2.5e+6}{\kilo\meter}, and reference noise spectral density as defined in the ESA Science Requirements Document~\cite{SciRD}. Unresolved galactic binaries' confusion noise does not affect SmBBHs considered here, and is therefore not included in the noise budget (although, to help guide the eye, it is plotted in Fig.~\ref{fig:waveforms}). The LISA response is modelled following the rigid adiabatic approximation (see, for example, Ref~\cite{2004PhRvD..69h2003R}). 
We set the mission duration to $\SI{1.26e8}{\second}$ and use a sampling cadence of $\SI{0.3}{\second}$. 

\textbf{Waveforms} -- In the LISA sensitivity window $\sim$  \SIrange{1e-4}{0.1}{\hertz}, SmBBHs are in their early inspiral. Current astrophysical understanding suggests that in the LISA band these sources could have non-zero eccentricity and the BHs spins have generic orientation with respect to the orbital angular momentum. We therefore employ \texttt{EFPE}~\cite{Klein:2021jtd}, a state-of-the-art frequency-domain waveform model that incorporates eccentricity and leading order relativistic spin-precession effects, with phasing accurate up to 3 post-Newtownian order.

\textbf{Parameter estimation} --
We perform fully Bayesian inference on simulated data, injecting the signal in zero noise data.
We use the \textsc{Balrog} codesuite, which provides a rigid adiabatic approximation response model for LISA~\cite{ak.ls.21}, interfaces to a number of stochastic samplers~\cite{john_veitch_2021_4470001,2021PhRvD.103j3006W,2020MNRAS.493.3132S,2013PASP..125..306F}, time- and frequency- domain waveforms for double white dwarfs and stellar mass binary black holes, infrastructure for time-delay-interferometry~\cite{2021LRR....24....1T}, and access to the \textsc{LDC} codebase under development within the LISA Consortium~\cite{ldc}.
Following closely the methodology in~\cite{balrog-ldc1}, we employ a Gaussian likelihood on the time delay interferometry (TDI) variables $A,E,T$, constructed as noise orthogonal combination of the Michelson TDI variables $X,Y,Z$. 
For each binary analysed, we generate data in frequency ranges broad enough to capture entirely the simulated signal. 
For the ``light'' (``heavy'') binaries, we set minimum and maximum frequencies to
$f_{\rm min}=\SI{0.0185}{\hertz}$ and $f_{\rm max}=\SI{0.1}{\hertz}$
($f_{\rm min}=\SI{0.0111}{\hertz}$ and $f_{\rm max}=\SI{0.1}{\hertz}$),
respectively. 
To reduce the computational cost, we evaluate inner products through a modified Clenshaw-Curtis integration scheme~\cite{clenshaw_method_1960}.
We sample parameters uniformly in their respective ranges (see Table~\ref{tab:space_based_priors} for details) the following parameters: 
chirp mass $\mathcal{M}_c$, 
dimensionless mass difference $\delta\mu = (m_1-m_2)/(m_1+m_2)$,
time to merger $t_m$,
squared eccentricity at \SI{0.01}{\hertz} $e^2_{0.01}$,
square roots of circularly polarized GW amplitudes $\sqrt{A_{L,R}}= \sqrt{A}(1\pm\cos \iota)$,
polarization angle $\psi$,
initial orbital phase $\phi_0$,
initial argument of periastron $\phi_e$,
ecliptic longitude $l$,
sine of ecliptic latitude $\sin b$,
dimensionless spin magnitudes $\chi_1,\chi_2$,
sine of spin latitudes $\sin b^{\chi}_1,\sin b^{\chi}_2$,
and spin longitudes $l^{\chi}_1,l^{\chi}_2$.

We use a nested sampling algorithm~\cite{2004aipc..735..395s} as implemented in \textsc{CPnest}, with 16000 livepoints multithreaded on $20$ cores for each of the four runs. Each run took approximately 3 weeks to reach its completion (with an evidence uncertainty of $\mathrm{d}Z=\pm0.1$), except for the heavy binary with high eccentricy, which required approximately $3$ months. In Table~\ref{tab:supp_parameters_heavy} and~\ref{tab:supp_parameters_light} parameters posterior median and 90\% confidence intervals are quoted, alongside the corresponding injected values.

%
%

\newcolumntype{Y}{>{\centering\arraybackslash}X}
\begin{table}[!ht]
\caption{ \label{tab:space_based_priors}
	Priors for the LISA parameter estimation.
} 
\centering
\renewcommand{\arraystretch}{1.4}
\begin{tabularx}{\textwidth}{@{}l | Y  Y | Y  Y @{}}
\hline \hline
Name & \multicolumn{2}{c|}{Light binary} & \multicolumn{2}{c}{Heavy binary} \\
\cline{1-5}
& High Ecc. & Low Ecc. & High Ecc. & Low Ecc. \\
\cline{2-5}
$\mathcal{M}_c~[M_\odot]$ 
& \multicolumn{2}{c|}{$\text{Uniform}(28.092,28.098)$} 
& \multicolumn{2}{c}{$\text{Uniform}(64.589,64.595)$} \\
$\delta\mu$ 
& \multicolumn{2}{c|}{$\text{Uniform}(0,0.9)$} 
& \multicolumn{2}{c}{$\text{Uniform}(0,0.9)$} \\
$t_m~[s]$ 
& \multicolumn{2}{c|}{$\text{Uniform}(-10^4,10^4)$} 
& \multicolumn{2}{c}{$\text{Uniform}(-10^4,10^4)$} \\
$e^2_{0.01}$ 
& $\text{Uniform}(0,3\times10^{-4})$ 
& $\text{Uniform}(0,3\times10^{-5})$ 
& $\text{Uniform}(0.00098,0.00102)$ 
& $\text{Uniform}(0,3 \times 10^{-5})$\\
$\sqrt{A_R}$ 
& \multicolumn{2}{c|}{$\text{Uniform}(0,1.158\times 10^{-4})$} 
& \multicolumn{2}{c}{$\text{Uniform}(0,6.1951 \times 10^{-5})$} \\
$\sqrt{A_L}$ 
& \multicolumn{2}{c|}{$\text{Uniform}(0,1.158\times 10^{-4})$}
& \multicolumn{2}{c}{$\text{Uniform}(0,6.1951 \times 10^{-5})$} \\
$\psi~[\text{rad}]$ 
& \multicolumn{2}{c|}{$\text{Uniform}(0,2 \pi)$}
& \multicolumn{2}{c}{$\text{Uniform}(0,2 \pi)$} \\
$\phi_0~[\text{rad}]$ 
& \multicolumn{2}{c|}{$\text{Uniform}(0,2 \pi)$} 
& \multicolumn{2}{c}{$\text{Uniform}(0,2 \pi)$} \\
$\phi_e~[\text{rad}]$ 
& \multicolumn{2}{c|}{$\text{Uniform}(0,2 \pi)$} 
& \multicolumn{2}{c}{$\text{Uniform}(0,2 \pi)$} \\
$l~[\text{rad}]$ 
& \multicolumn{2}{c|}{$\text{Uniform}(4,6)$} 
& \multicolumn{2}{c}{$\text{Uniform}(5.4,5.7)$} \\
$\sin b$ 
& \multicolumn{2}{c|}{$\text{Uniform}(-0.999,-0.993)$} 
& \multicolumn{2}{c}{$\text{Uniform}(-0.84,-0.82)$} \\
$\chi_{1},\chi_2$ 
& \multicolumn{2}{c|}{$\text{Uniform}(0,1)$}
& \multicolumn{2}{c}{$\text{Uniform}(0,1)$}\\
$l^\chi_1,l^\chi_2~[\text{rad}]$ 
& \multicolumn{2}{c|}{$\text{Uniform}(0,2 \pi)$} 
& \multicolumn{2}{c}{$\text{Uniform}(0,2 \pi)$} \\
$\sin b^\chi_1, \sin b^\chi_2$ 
& \multicolumn{2}{c|}{$\text{Uniform}(-1,1)$} 
& \multicolumn{2}{c}{$\text{Uniform}(-1,1)$} \\
\hline \hline
\end{tabularx}
\end{table}

\begin{table}[ht]
\caption{ \label{tab:ground_based_priors}
	Priors for the ground-based parameter estimation. Note that whilst we use uniform priors in the component masses, following \cite{Veitch:2014wba} we sample in $\lbrace \mathcal{M}, q \rbrace$ as these variables are less correlated.
} 
\centering
\renewcommand{\arraystretch}{1.5}
\begin{tabularx}{0.5\textwidth}{>{\raggedright}p{7em} | c | c }
\hline \hline
  Name \quad\quad\quad \phantom{.} & Light & Heavy \\
  \hline
  $m_1 \, [M_{\odot}]$ & $\text{Uniform}(20,50)$ & $\text{Uniform}(70,100)$ \\
  $m_2 \, [M_{\odot}]$ & $\text{Uniform}(20,50)$ & $\text{Uniform}(50,80)$ \\
  $\chi_1, \chi_2$ & $\text{Uniform}(0,0.99)$ & $\text{Uniform}(0,0.99)$ \\
  $\theta_1, \theta_2 \, [\rm rad]$ & $\text{Sine}(0,\pi)$ & $\text{Sine}(0,\pi)$ \\
  $\phi_{12}, \phi_{\rm JL} \,[\rm rad]$ & $\text{Uniform}(0,2 \pi)$ & $\text{Uniform}(0,2 \pi)$ \\
  $\theta_{\rm JN}\, [\rm rad]$ & $\text{Sine}(0,\pi)$ & $\text{Sine}(0,\pi)$ \\
  $\psi$\, [\rm rad] & $\text{Uniform}(0,\pi)$ & $\text{Uniform}(0,\pi)$ \\
  $\alpha$\, [\rm rad] & $\text{Uniform}(0,2\pi)$ & $\text{Uniform}(0,2 \pi)$ \\
  $\delta\, [\rm rad]$ & $\;\text{Cosine}(-\pi/2,\pi/2)\;$ & $\;\text{Cosine}(-\pi/2,\pi/2)\;$ \\
  \hline \hline
\end{tabularx}
\end{table}

\section{Ground-based data analysis}\label{supp:3G}
\textbf{Detectors} -- For the ground-based analysis, we consider the A+ detector network consisting of LIGO-Hanford, LIGO-Livingston and Virgo (HLV+), the triangular ET-D configuration \cite{2020JCAP...03..050M} placed at the current Virgo site, or a single L-shaped CE detector \cite{Reitze:2019iox} in the location of LIGO-Hanford. For Hanford and Livingston, we adopt the A+ design target for the fifth observing run \cite{psdobs,KAGRA:2013rdx}, and for Virgo we use its high-limit target sensitivity \cite{psdobs,KAGRA:2013rdx}. For the 3G detectors, we use the ET-D and Cosmic Explorer noise curves taken from \cite{psd3g}.

\textbf{Waveforms} -- By the time the signals considered here reach a GW frequency of $10\,\mathrm{Hz}$, the orbit will have circularised~\cite{Peters:1964zz} and hence we use a state-of-the-art waveform model for the inspiral, merger and ringdown of quasi-circular binaries, \texttt{IMRPhenomXPHM}~\cite{Pratten:2020fqn,Garcia-Quiros:2020qpx,Pratten:2020ceb}. Following the general framework to build waveform models that include general-relativistic spin-precession effects \cite{Schmidt:2010it,Schmidt:2012rh}, the model applies a time-dependent rotation that encapsulates the precession of the orbital plane to a set of approximate aligned-spin $h_{\ell m}$-modes. The approximate aligned-spin modes calibrated against numerical relativity simulations in \texttt{IMRPhenomXPHM} are the $(\ell,|m|) = \lbrace (2,2),(2,1),(3,3),(3,2),(4,4)\rbrace$ modes. 

\textbf{Parameter estimation} -- We perform Bayesian inference utilising the \textsc{Bilby} inference library~\cite{Ashton:2018jfp} in combination with the nested sampler \textsc{dynesty}~\cite{2020MNRAS.493.3132S} to sample the BBH parameter space and infer the posterior densities. In order to help improve convergence of the posteriors, we use $\sim \mathcal{O}(\text{few} \times 10^3)$ live points and adopt a stopping criterion on the evidence $Z$ of $\Delta \ln Z = 0.1$. Due to the very large SNRs of the injected signals, the nested sampling algorithm is extremely slow to converge and more efficient sampling methods, such as slice sampling, offer significant computational improvement over the default random-walk proposals. Each run took $\sim 2-3$ weeks on $32$ cores. The simulated signals are injected into zero noise, which should be broadly equivalent to the results obtained by averaging over a large number of injections into different realizations of Gaussian noise. The respective PSD enters only in the likelihood calculation. For both signals, we analyse data segments of $8\,\mathrm{s}$ duration with a sampling rate of $1024\,\mathrm{Hz}$. We sample in the (detector-frame) chirp mass $\mathcal{M}_c$ and mass ratio $q$ but adopt priors that are uniform in the component masses \cite{Veitch:2014wba}. The complete set of priors is given in Table~\ref{tab:ground_based_priors}. We assume the binaries are observed at a Global Positioning System (GPS) reference time of \texttt{1706962960.0} at the geocenter.

\section{Detailed results for all analyses}\label{supp:plots}

In the main text, we have presented the overall results, presenting some of the key illustrative plots. Here we show detailed results for all the analyses that we have carried out in the context of this study.

A top level summary of the symmetric 90\% credible intervals of the parameters recovered in all the analyses -- LISA and the different ground-based observatories, HLV+, ET and CE -- for all the sources is given in Tables~\ref{tab:supp_parameters_heavy} and~\ref{tab:supp_parameters_light}. 

\subsection{Summary plots for the heavy binary}

\begin{figure}[tbp]
	\centering
	\includegraphics[width=0.45\textwidth]{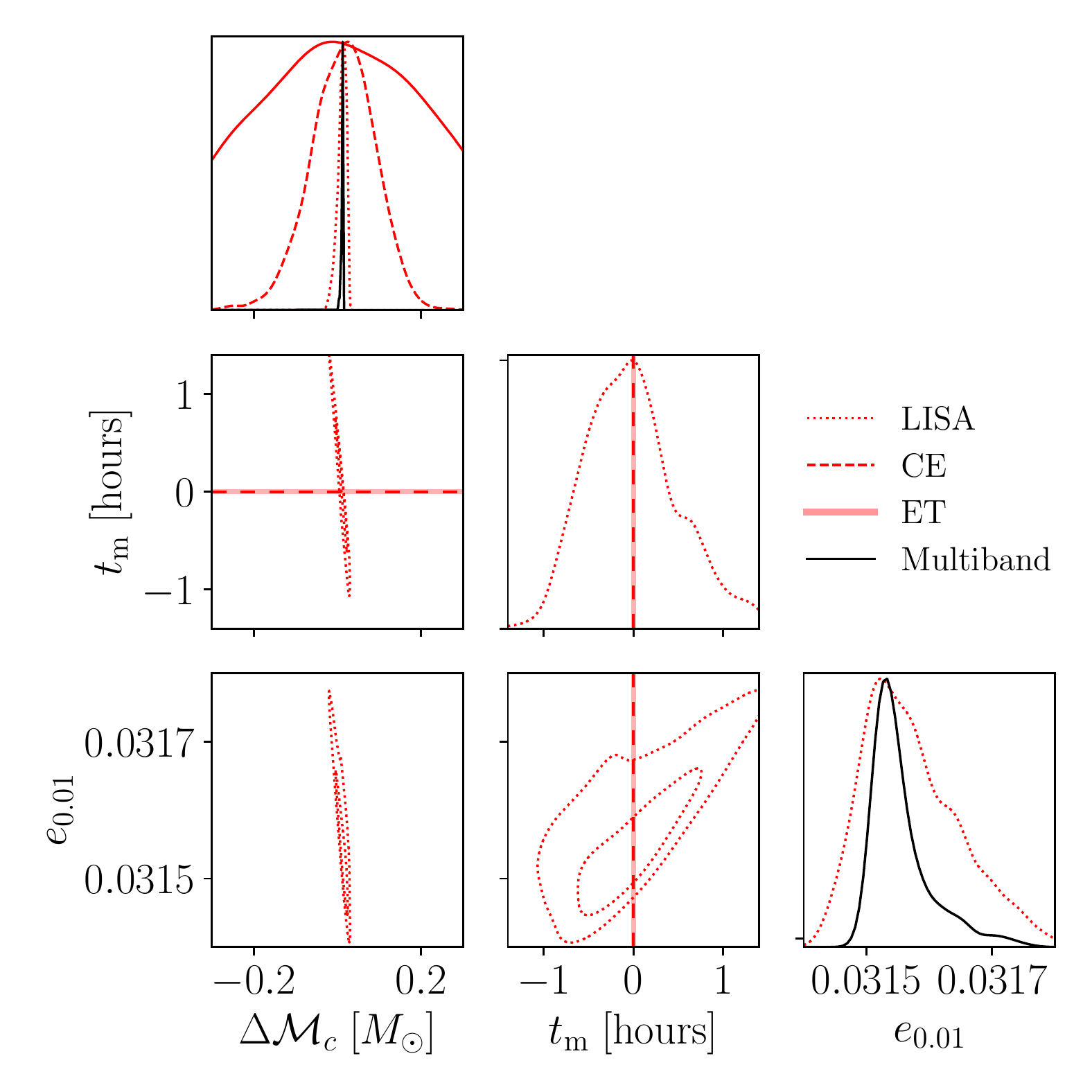}
    \includegraphics[width=0.45\textwidth]{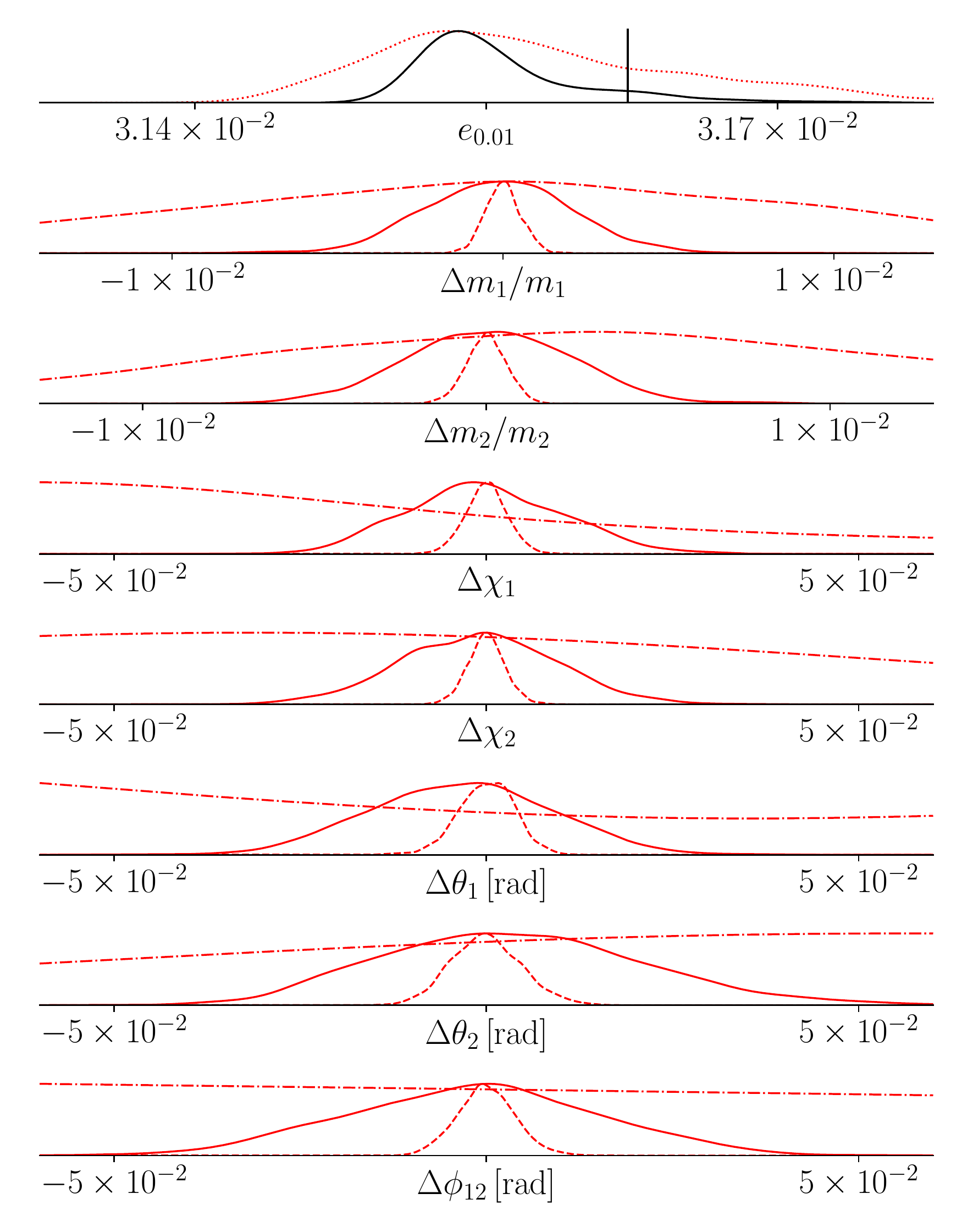}
	\caption{ \label{fig:supp_Mc_tm_e0}
	    Selected posterior density functions for the ``heavy'' ($85\,M_\odot-65\,M_\odot$),  high-eccentricity ($e_{0.01} = 0.0316$) source, same as Figs.~\ref{fig:Mc_e0_tm} and~\ref{fig:intrinsic}, which refer to the ``light'' binary. Left: LISA posteriors (dotted red) on the chirp mass $\mathcal{M}_c$, time to merger $t_m$, and eccentricity $e_{0.01}$. 
		Solid/dashed red lines show posteriors from third generation ET/CE ground-based instruments.
	    The LISA $\mathcal{M}_c$ measurement is strongly degenerate with $t_m$ (and, to a lesser extent, with $e_{0.01}$) which is broken when combining with the merger time measured from any ground-based instrument leading to significant improvements in the combined, multiband measurements (shown in black). Right: One-dimensional marginalised posteriors on the intrinsic parameters:
		$10\,\mathrm{mHz}$ eccentricity $\ecc$, 
		detector frame component masses $m_{1(2)}$, 
		dimensionless spin magnitudes $\chi_{1(2)}$,
		spin tilt angles $\theta_{1(2)}$, 
		and the angle between the in-plane spins $\phi_{12}$.
		Dot/dash/solid/dot-dash lines indicate LISA/CE/ET/HLV+ posteriors. 
		LISA measures eccentricity extremely well from the early inspiral (the black curve shows the improvement with the multiband measurement of $t_m$) and the 3G instruments measure the other intrinsic parameters with exquisite accuracy.
	    The injected eccentricity is indicated; for all other parameters, $\Delta$ denotes a shift from the injected value. 
	    }
\end{figure}

In the main text we have shown plots based on the results for the light binary for the chirp mass, time to merger and eccentricity (Fig.~\ref{fig:Mc_e0_tm}), and intrinsic (Fig.~\ref{fig:intrinsic}) parameters. Here we show the same plots, but for the heavy binary. 
Figure~\ref{fig:supp_Mc_tm_e0} (right panel) shows the posterior distribution on the chirp mass, merger time and eccentricity (right panel) for the heavy source, equivalent to Fig.~\ref{fig:supp_Mc_tm_e0} in the main text. The right panel of Figure~\ref{fig:supp_Mc_tm_e0} shows the marginalised posterior distributions of the physical parameters of the heavy source, the equivalent of Fig.~\ref{fig:intrinsic} in the main text.

%
%

\subsection{Individual mass measurements}

\begin{figure*}[tbp]
	\centering
    \includegraphics[width=0.9\textwidth]{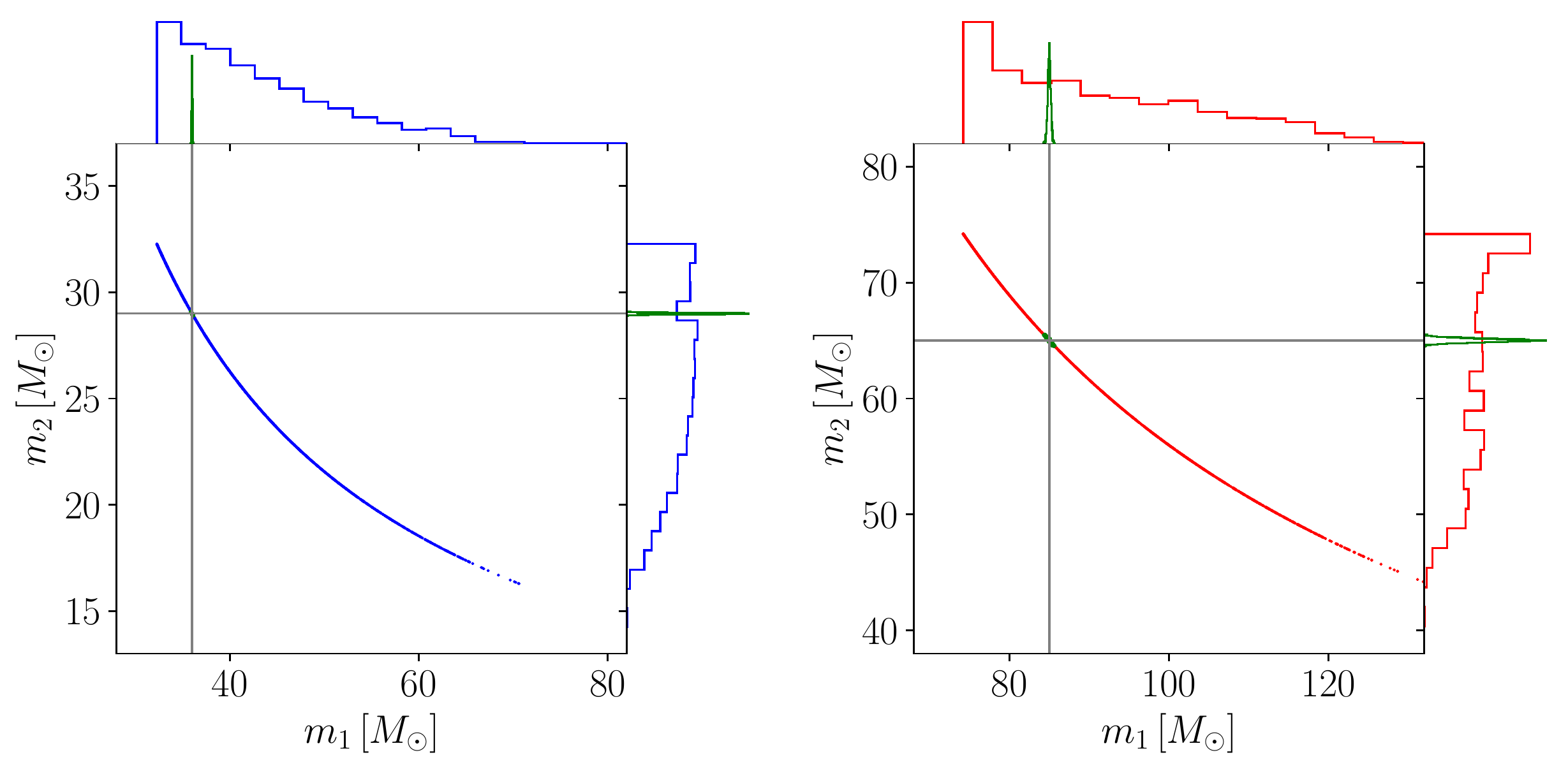}
	\caption{ \label{fig:masses_LISA_ET}
	    Marginalised posterior density functions on the component masses of the two binary systems considered in this study obtained in observations with different instruments.
		\emph{Left}: posteriors for the light ($36\,M_\odot-29\,M_\odot$),  high-eccentricity ($e_{0.01} = 0.007$) binary recovered with LISA (blue) and ET (green) observations. \emph{Right}: posteriors for the heavy ($85\,M_\odot-65\,M_\odot$) high-eccentricity ($e_{0.01} = 0.0316$) binary obtained with LISA (red) and ET (green). 
	}
\end{figure*}

In Fig.~\ref{fig:masses_LISA_ET} we show the posterior density functions on the individual masses, $m_1$ and $m_2$, for LISA and ET observations. ET is meant to represent the quality of observations for a 3G ground-based observatory operating in the LISA time-frame. Note that results obtained with CE are comparable, but actually better by a factor $\approx 2$, see Tables~\ref{tab:supp_parameters_heavy},~\ref{tab:supp_parameters_light} and Figs.~\ref{fig:corner_ground_GW150914} and~\ref{fig:corner_ground_GW190521}

%
%

\subsection{Measuring the eccentricity}

\begin{figure}[tbp]
	\centering
	\includegraphics[width=0.6\textwidth]{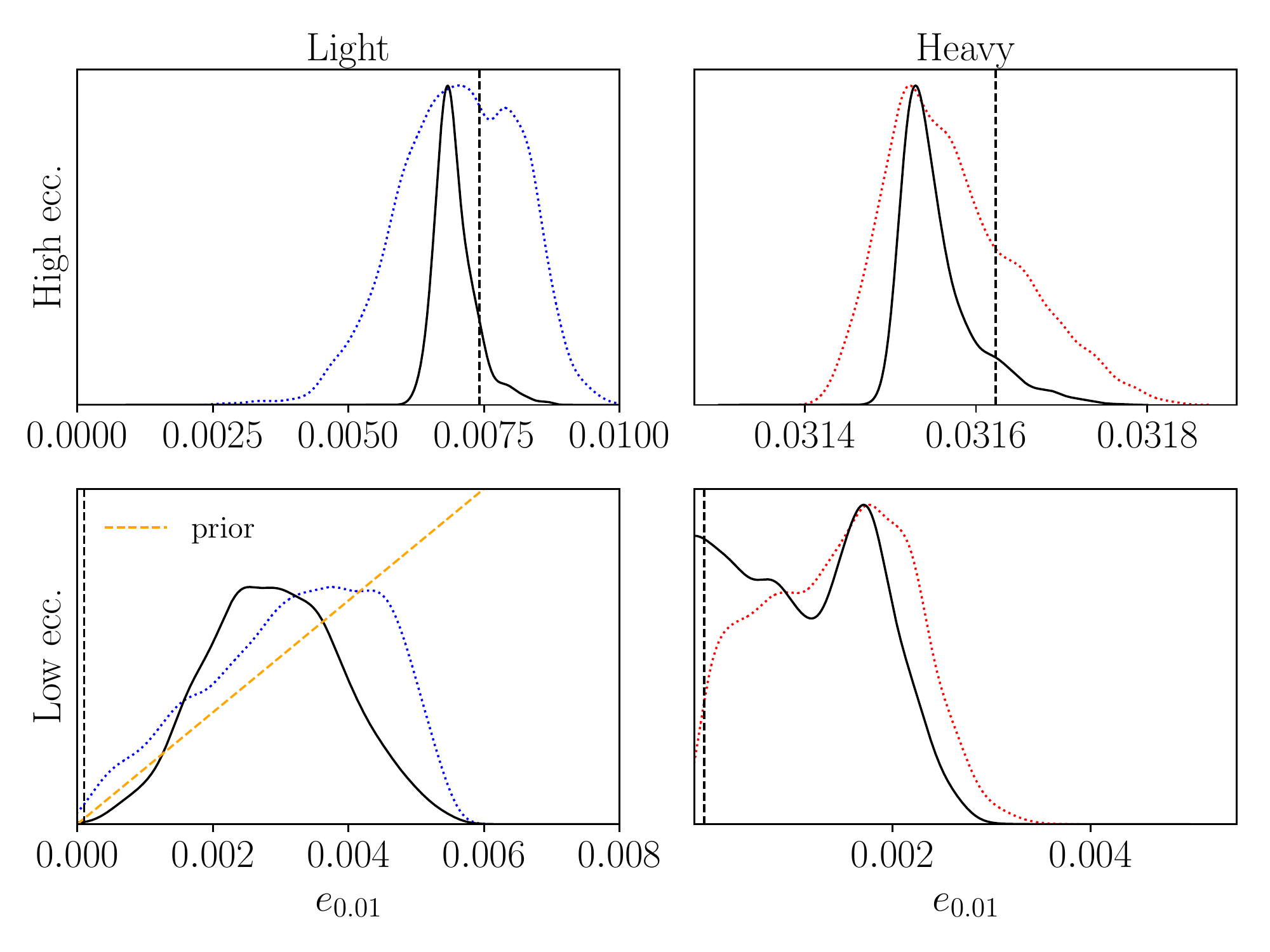}
	\caption{ \label{fig:e0}
	Marginalised posterior density functions on the eccentricity for the light ($36\,M_\odot-29\,M_\odot$) binary system (plots on the left) and the heavy ($85\,M_\odot-65\,M_\odot$) binary system (plots on the right) with high (top row) and low (bottom row) eccentricity (see Tables~\ref{tab:summary_parameters} and~\ref{tab:supp_parameters_heavy} and~\ref{tab:supp_parameters_light} for the parameter values, also shown as black dashed lines). Dotted lines are the posteriors recovered with LISA observations only (blue: light binary; red: heavy binary). The multi-band posterior density function, \textit{i.e.} the result obtained combining both LISA and ground-based observations (here we use specifically the results obtained with ET) are shown by the solid black line. 
	Also show for reference in the bottom left panel is the prior distribution.
	}
\end{figure}

While regardless of the eccentricity of the binary at formation, we expect that the orbit has circularised due to radiation reaction by the time the signal enters the band of ground-based observatories, some residual eccentricity may well be present in the LISA sensitivity window. For this reason, for both the heavy and light binary we consider two different eccentricity values, labelled high and low (see Tables~\ref{tab:summary_parameters} and~\ref{tab:supp_parameters_heavy}). The results concerning the measurement of the eccentricity for both LISA-only and multi-band observations are shown in Fig.~\ref{fig:e0}. They clearly show that for the ``light'' (``heavy'') binary an eccentricity of $7\times 10^{-3}$ ($3\times 10^{-2}$) can be measured and distinguished from zero at high confidence, well above $99\%$ confidence. Conversely, for an eccentricity of $1\times 10^{-4}$ one can only provide an upper-limit to the eccentricity of the binary. These values can provide a guide as of what is the level of eccentricity for SmBBHs that will be measurable with LISA, and the one for which only upper-limits can be set. We need however to caution the reader that the exact values will depend on a source-by-source basis, as they are affected by SNR, signal duration in band, values of masses and spins. 

%
%

\subsection{(Almost) full corner plots}

The binaries considered in this study are described by the full set of 17 parameters (15 in the context of ground-based observations, where we assume the orbit has circularised below the limit of detection of eccentricity, and we therefore consider exactly circular binaries). Full corner plots would therefore be illegible, and we are not presenting them here. We show however corner plots for the most of the parameters of astrophysical interest below for the full set of runs.

\begin{figure}
	\centering
	\includegraphics[width=\textwidth]{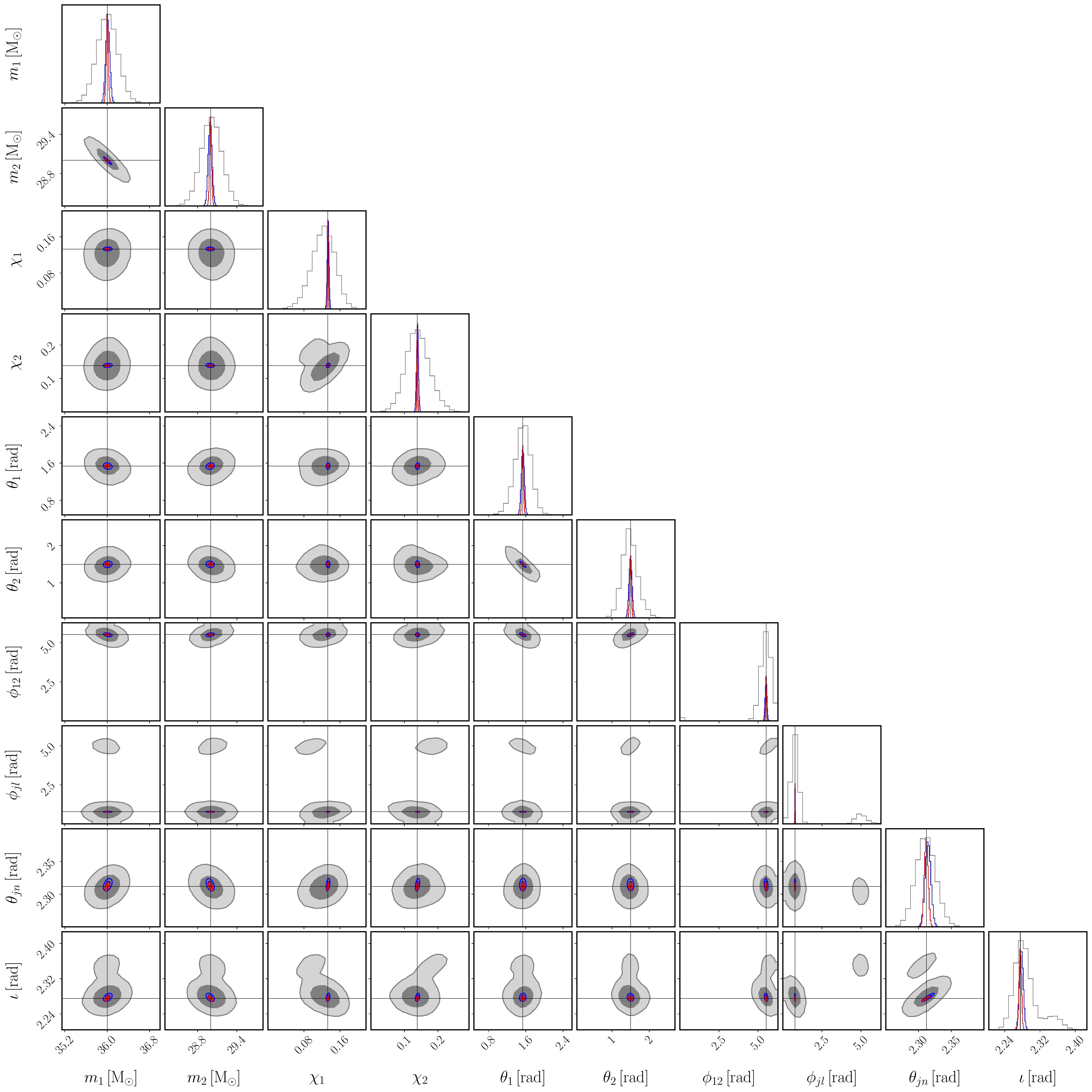}
	\caption{ \label{fig:corner_ground_GW150914}
	Marginalised posterior density functions on the mass and spin parameters (details are provided in Table~\ref{tab:supp_parameters_light}) for observations of the light ($36\,M_\odot-29\,M_\odot$) binary system with different ground-based observatories: HLV+ (grey contours: 50\% and 90\% probability regions) ET (blue) CE (red) and O5 (grey). 
	}
\end{figure}

\begin{figure}
	\centering
	\includegraphics[width=\textwidth]{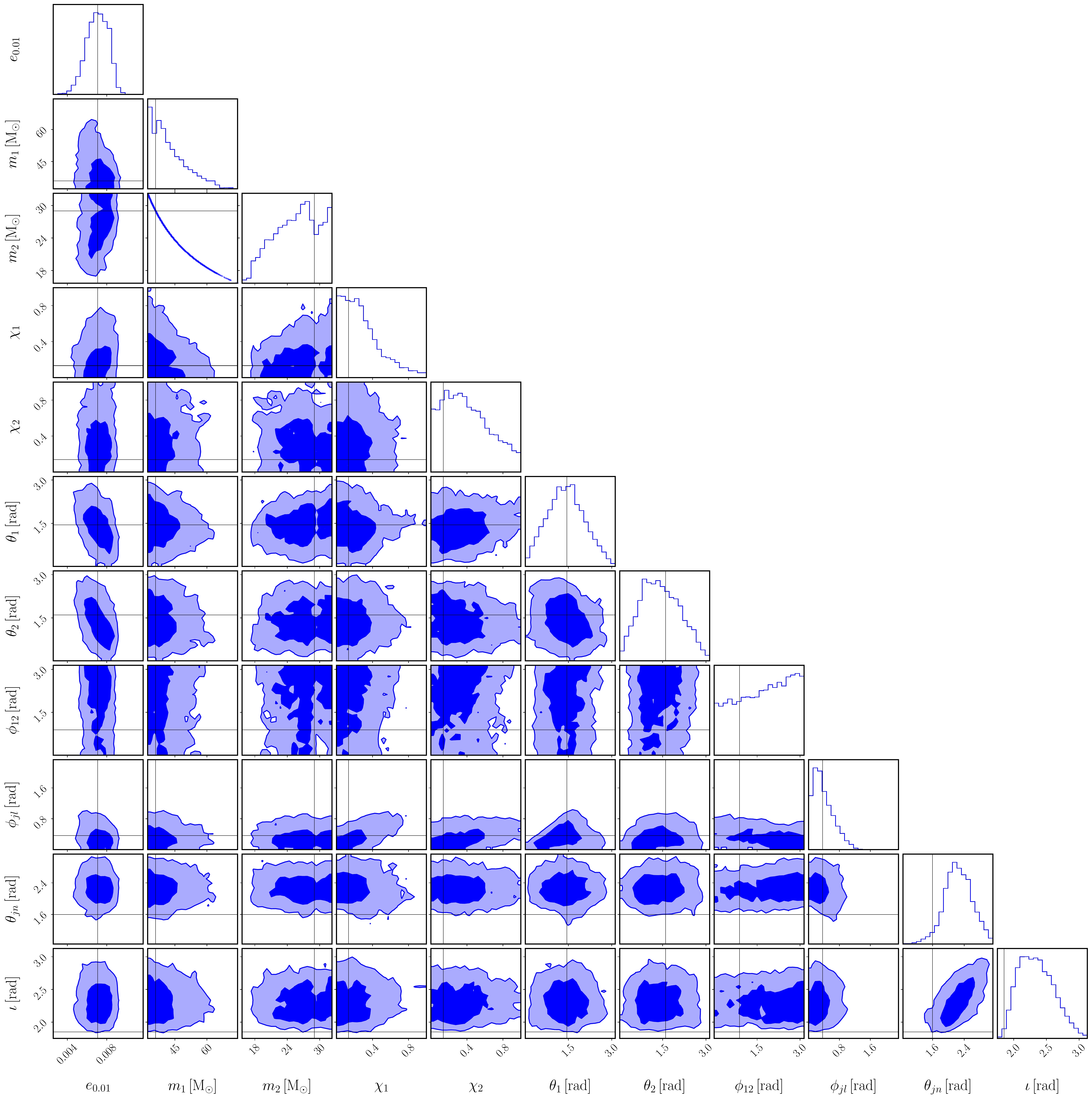}
	\caption{ \label{fig:corner_LISA_GW150914_hiecc}
	Marginalised posterior density functions on the eccentricity, mass and spin parameters (details are provided in Table~\ref{tab:supp_parameters_light}) for observations of the light ($36\,M_\odot-29\,M_\odot$), high eccentricity ($e_{0.01} = 0.0077$) binary system with LISA. The dark (light) blue areas in the two-dimensional plots refer to the 50\% (90\%) posterior probability regions. 
	}
\end{figure}

\begin{figure}
	\centering
	\includegraphics[width=\textwidth]{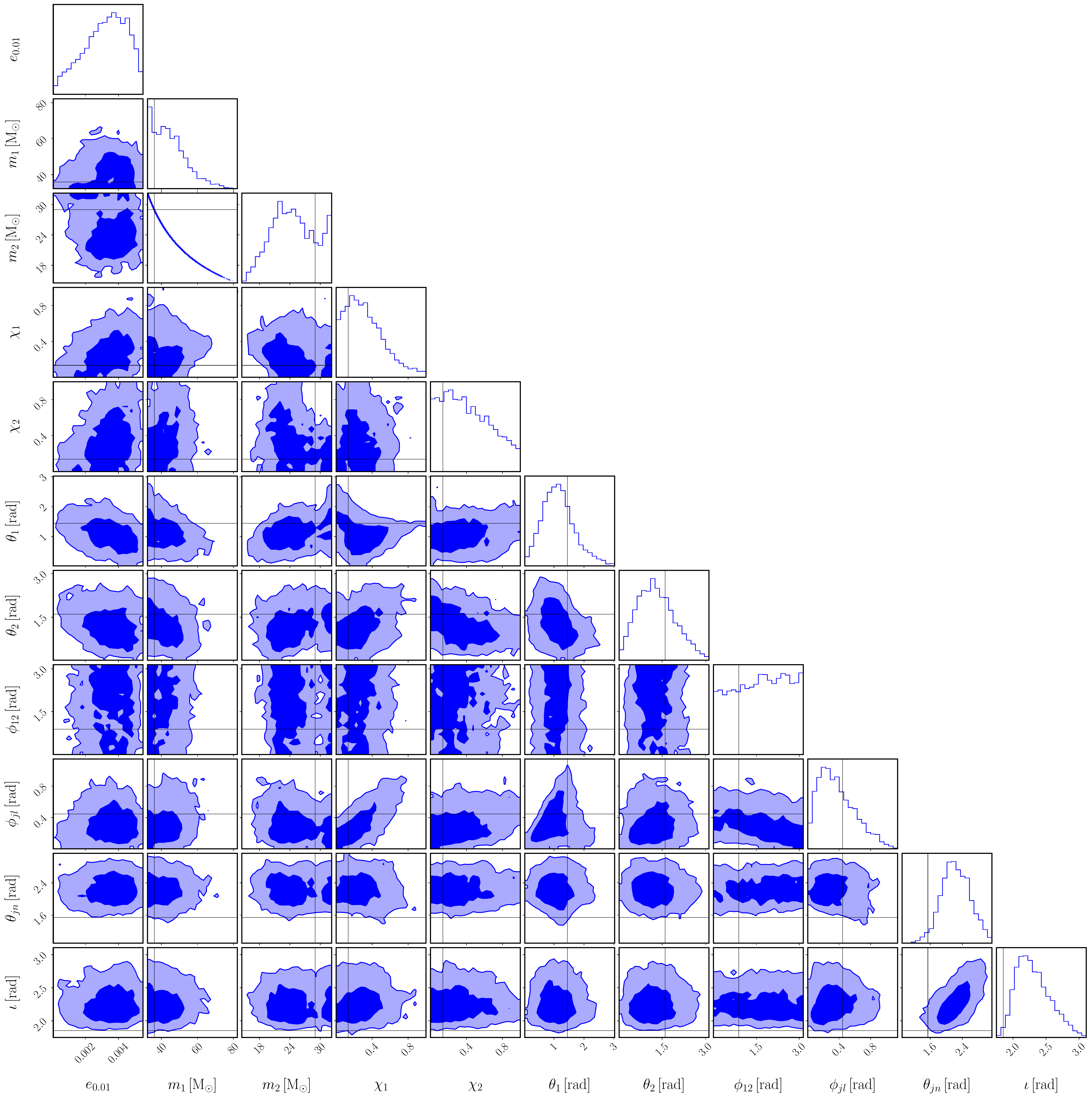}
	\caption{ \label{fig:corner_LISA_GW150914_loecc}
	Marginalised posterior density functions on the eccentricity, mass and spin parameters (details are provided in Table~\ref{tab:supp_parameters_light}) for observations of the light ($36\,M_\odot-29\,M_\odot$), low eccentricity ($e_{0.01} = 1\times 10^{-4}$) binary system with LISA. The dark (light) blue areas in the two-dimensional plots refer to the 50\% (90\%) posterior probability regions.
	}
\end{figure}

Fig~\ref{fig:corner_ground_GW150914} show results for the physical parameters of the light source for the three different configurations of ground-based observatories: ET, CE and HLV+. The equivalent corner plots for the same source in the case of LISA observations are Fig~\ref{fig:corner_LISA_GW150914_hiecc} (for the high-eccentricity case) and Fig~\ref{fig:corner_LISA_GW150914_loecc} (for the low-eccentricity case). 90\% credible intervals are summarised in Table~\ref{tab:supp_parameters_light}.

\begin{figure}
	\centering
	\includegraphics[width=\textwidth]{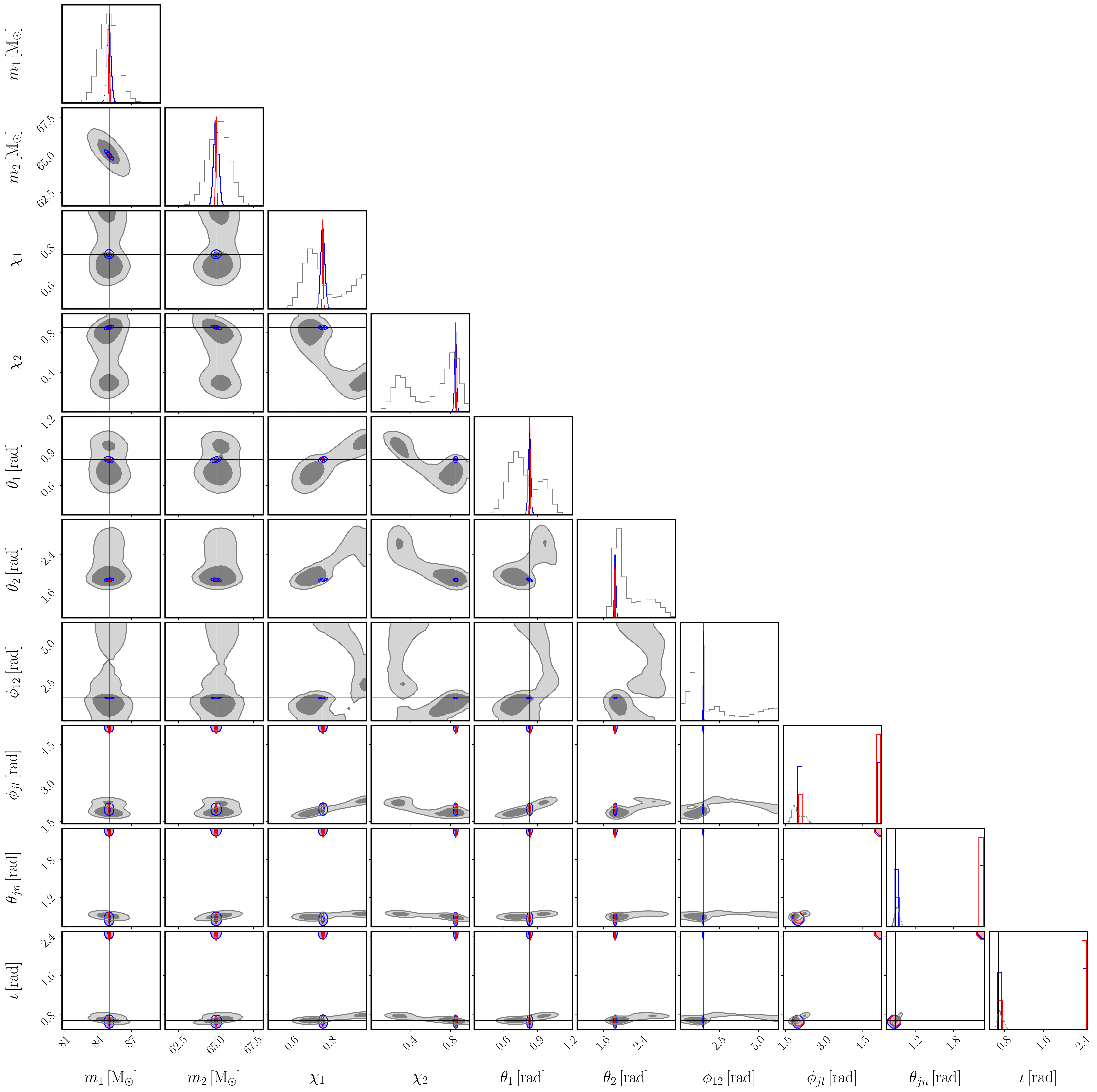}
	\caption{ \label{fig:corner_ground_GW190521}
	Marginalised posterior density functions on the mass and spin parameters (details are provided in Table~\ref{tab:supp_parameters_heavy}) for observations of the heavy ($85\,M_\odot-65\,M_\odot$) binary system with different ground-based observatories: HLV+ (grey contours: 50\% and 90\% probability regions) ET (blue) CE (red) and O5 (grey). 
	}
\end{figure}

\begin{figure}
	\centering
	\includegraphics[width=\textwidth]{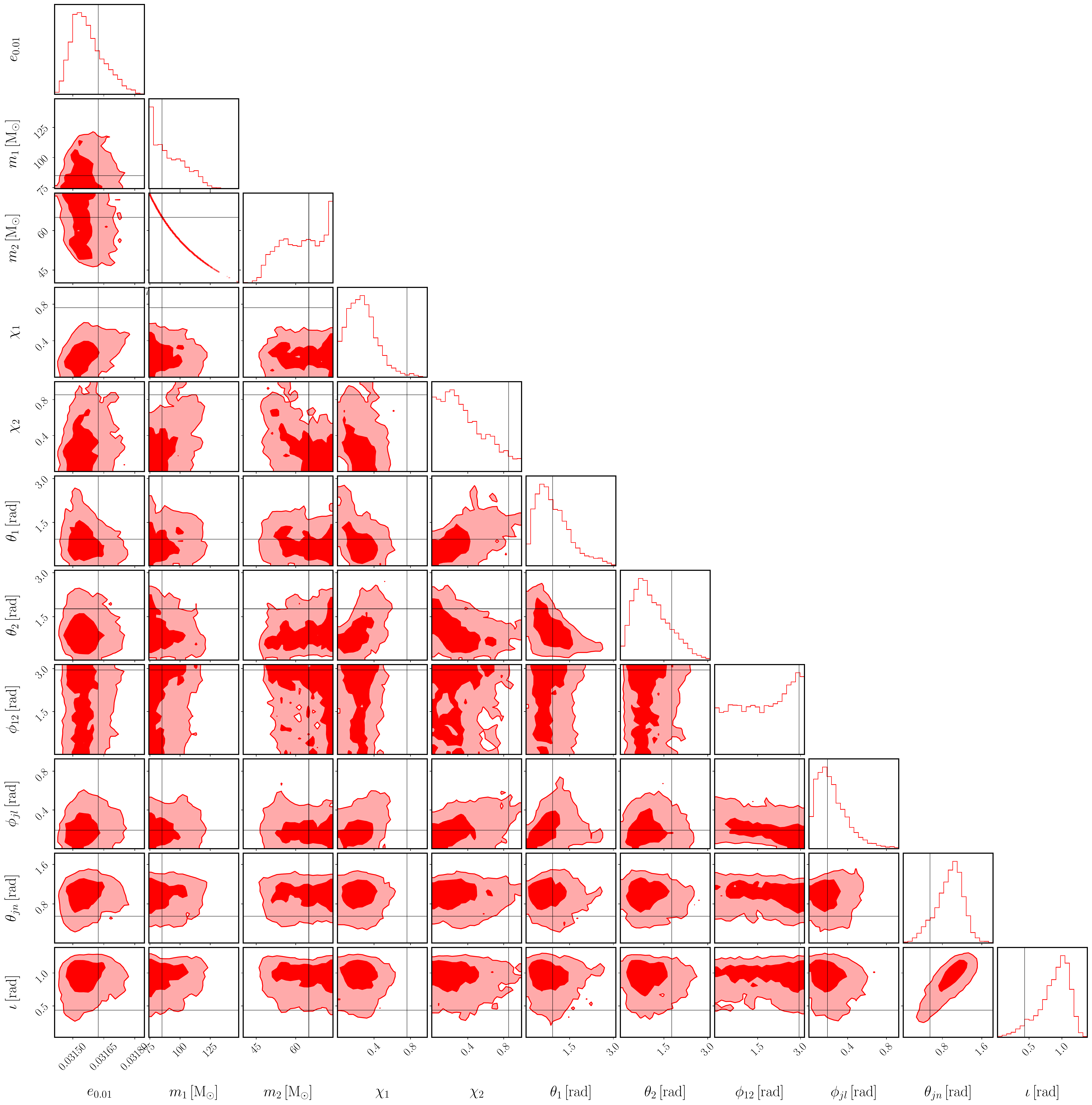}
	\caption{ \label{fig:corner_LISA_GW190521_hiecc}
	Marginalised posterior density functions on the eccentricity, mass and spin parameters (details are provided in Table~\ref{tab:supp_parameters_heavy}) for observations of the heavy ($85\,M_\odot-65\,M_\odot$), high eccentricity ($e_{0.01} = 0.0316$) binary system with LISA. The dark (light) red areas in the two-dimensional plots refer to the 50\% (90\%) posterior probability regions. 
	}
\end{figure}

\begin{figure}
	\centering
	\includegraphics[width=\textwidth]{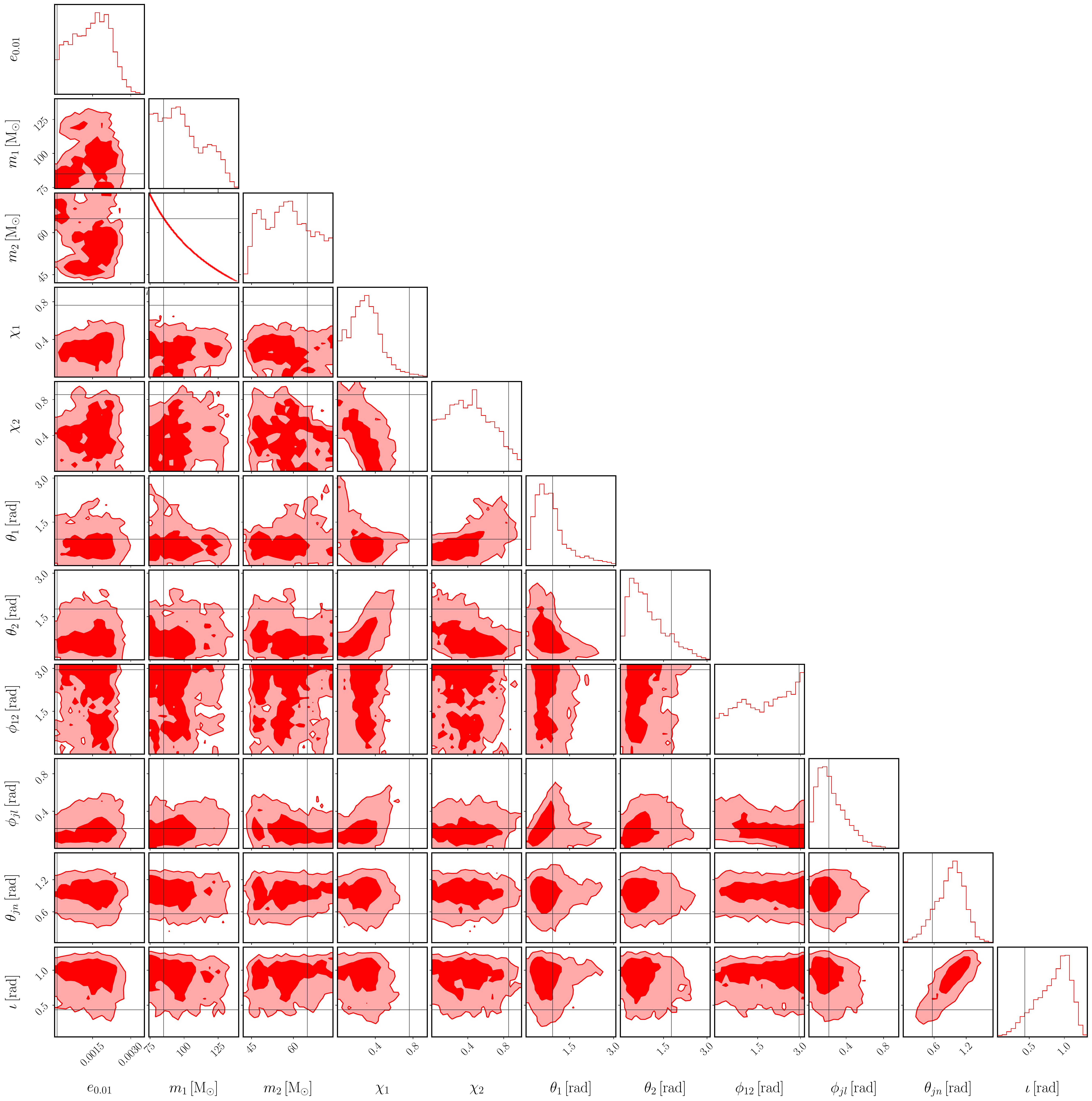}
	\caption{ \label{fig:corner_LISA_GW190521_loecc}
	Marginalised posterior density functions on the eccentricity, mass and spin parameters (details are provided in Table~\ref{tab:supp_parameters_heavy}) for observations of the heavy ($85\,M_\odot-65\,M_\odot$), low eccentricity ($e_{0.01} = 1\times 10^{-4}$) binary system with LISA. The dark (light) red areas in the two-dimensional plots refer to the 50\% (90\%) posterior probability regions. 
	}
\end{figure}

Fig~\ref{fig:corner_ground_GW190521} show results for the physical parameters of the heavy source for the three different configurations of ground-based observatories: ET, CE and HLV+. The equivalent corner plots for the same source in the case of LISA observations are Fig~\ref{fig:corner_LISA_GW190521_hiecc} (for the high-eccentricity case) and Fig~\ref{fig:corner_LISA_GW190521_loecc} (for the low-eccentricity case). 90\% credible intervals are summarised in Table~\ref{tab:supp_parameters_heavy}.

\newpage

\newpage

\newpage

\newpage

\newpage

\newpage

\newpage

\end{document}